\documentclass{ws-jai-preprint}
\usepackage[flushleft]{threeparttable}

\usepackage{graphicx}	
\usepackage{amsmath}	
\usepackage{subfiles}
\usepackage{subcaption}
\usepackage[export]{adjustbox}
\usepackage{tabularx}
\usepackage{siunitx}
\usepackage{color,soul}
\usepackage{pifont}
\usepackage{booktabs}
\usepackage[hidelinks]{hyperref}

\begin{document}

\catchline{}{}{}{}{} 

\markboth{G. V. C. Allen, S. Pegwal, D. I. L. de Villiers et al.}{Circuit Modeling for In Situ 21 cm Radiometer Calibration}

\title{Circuit Modeling for In Situ 21 cm Radiometer Calibration}

\author{G. V. C. Allen$^{1^{*\dagger}}$,
S. Pegwal$^{1\dagger}$,
D. I. L. de Villiers$^{1\dagger}$,
D. Anstey$^{2,3}$,
K. Artuc$^{2,3}$,
H. T. J. Bevins$^{2,3}$,
G. Bernardi$^{7}$,
M. Bucher$^{1,6}$,
S. Carey$^{2}$,
J. Cavillot$^{8}$,
R. Chiello$^{9}$,
A. S. Chu$^{2,3}$,
W. Croukamp$^{1}$,
J. Cumner$^{2,3}$,
A. K. Dash$^{2,3}$,
S. Dasgupta$^{2,4}$,
E. de Lera Acedo$^{2,3}$,
J. Dhandha$^{3,4}$,
A. Dragovic$^{2,3}$,
J. A. Ely$^{2}$,
A. Fialkov$^{3,4}$,
T. Gessey-Jones$^{2,3}$,
W. J. Handley$^{3,4}$,
C. Kirkham$^{2,3}$,
G. Kulkarny$^{10}$,
S. A. K. Leeney$^{2,3}$,
A. Magro$^{5}$,
P. Daan	Meerburg$^{11}$,
S. Mittal$^{2,3}$,
D. Molnar$^{2,3}$,
R. S. Patel$^{2,3}$,
J. H. N. Pattison$^{2,3}$,
C. M. Pieterse$^{1}$,
J. R. Pritchard$^{12}$,
G. Rajpoot$^{2,3}$,
N. Razavi-Ghods$^{2}$,
D. Robins$^{2,3}$,
I. L. V. Roque$^{2}$,
A. Saxena$^{2,3}$,
K. H. Scheutwinkel$^{2,3}$,
P. Scott$^{2}$,
E. Shen$^{2,3}$,
P. H. Sims$^{2,3}$,
M. Spinelli$^{13,14}$,
J. Zhu$^{2,15}$
}

\address{
$^{1}$Department of Electrical and Electronic Engineering, Stellenbosch University, Stellenbosch, 7602, South Africa.\\
$^{2}$Astrophysics Group, Cavendish Laboratory, University of Cambridge, J. J. Thomson Avenue, Cambridge, CB3 0HE, UK.\\
$^{3}$Kavli Institute for Cosmology in Cambridge, University of Cambridge, Madingley Road, Cambridge, CB3 0HA, UK.\\
$^{4}$Institute of Astronomy, University of Cambridge, Madingley Road, Cambridge, CB3 0HA, UK.\\
$^{5}$Institute of Space Sciences and Astronomy, University of Malta, Msida, Malta, MSD 2080, Malta.\\
$^{6}$Laboratoire AstroParticule et Cosmologie, Université Paris-Cité, 10 Rue Alice Domon et Léonie Duquet, Paris, 75013, France.\\
$^{7}$INAF-Istituto di Radio Astronomia, Via Gobetti 101, Bologna, 40129, Italy.\\
$^{8}$Antenna Group, Université catholique de Louvain, Louvain-la-Neuve, 1348, Belgium.\\
$^{9}$Physics Department, University of Oxford, Parks Road, Oxford, OX1 3PU, UK.\\
$^{10}$Department of Theoretical Physics, Tata Institute of Fundamental Research, Homi Bhabha Road, Mumbai, 400005, India.\\
$^{11}$Faculty of Science and Engineering, University of Groningen, Nijenborgh 4, Groningen, 9747 AG, Netherlands.\\
$^{12}$Max Planck Institute for Radio Astronomy, Auf dem Hügel 69, 53121 Bonn, Germany\\
$^{13}$Observatoire de la Côte d’Azur, Nice, France.\\
$^{14}$Department of Physics and Astronomy, University of the Western Cape, Robert Sobukhwe Road, Bellville, 7535, South Africa.\\
$^{15}$National Astronomical Observatory, Chinese Academy of Science, Beijing, 100101, China.
}

\maketitle

\corres{$^{*}$Corresponding author. E-mail: gvcallen@gmail.com \\ $^{\dagger}$Primary authors.}

\begin{history}
\end{history}

\begin{abstract}
Recent experiments in cosmology, particularly those aimed at detecting the faint, redshifted, global 21 cm hydrogen line (depth $\lesssim \SI{200}{mK}$, $z > 7.5$), have imposed stringent new requirements on radiometer calibration. In this work, we present a framework for circuit modeling and parameter inference to strengthen these calibration pipelines. This new approach enables in situ characterization of otherwise immeasurable systematics using physically motivated models. A combination of frequentist and Bayesian techniques are employed in a pipeline that supports iterative modeling, robust parameter estimation, and detailed uncertainty quantification. The framework is applied to the REACH telescope, where the precise correction of variations in the radio signal paths arising from component aging or environmental effects is critical. Circuit models of REACH's calibration sources are developed, with the goal of predicting source temperature corrections that are conventionally obtained from laboratory measurements. By fitting the models to measured data using a convolutional cost function, a strong agreement with RMS residuals no worse than $\SI{-37}{dB}$ is obtained. However, Bayesian inference reveals that the resulting temperature corrections can have uncertainties on the order of $1$ to $\SI{2}{K}$, caused by reflection coefficient degeneracies, measurement noise, and errors in the models. To combat this, posteriors obtained from laboratory measurements are employed as updated priors, reducing correction uncertainties down to $\SI{75}{mK}$. Ultimately, the framework provides a means of dynamically accounting for drift in system non-idealities over time, addressing the increasing precision demands of global 21 cm radio astronomy.
\end{abstract}


\newpage
\section{Introduction}

Over the last century, our understanding of the universe has advanced dramatically, driven by major developments in observational techniques and theoretical models. Experiments across the electromagnetic spectrum have revealed the presence of galaxies, clusters, and large-scale cosmic structures, allowing us to map the universe over time. However, a critical period in cosmic history, the first billion years after the Big Bang, remains poorly understood. This era, known as the Dark Ages, Cosmic Dawn, and Epoch of Reionization, marks the time when the first stars and galaxies formed, initiating the processes of cosmic heating, reionization, and structure formation. Studying this period is essential for understanding how the complex universe we observe today came into existence.

Optical and near-infrared observations are limited in their ability to probe the cosmic dawn. During this epoch, the earliest luminous sources were still forming, and the universe was largely filled with neutral hydrogen, which obscured the light from emerging stars. An alternative approach is offered by low-frequency radio astronomy, which is theoretically capable of detecting the redshifted 21 cm hyperfine transition line emitted by neutral hydrogen atoms. This 21 cm signal is expected to be observable as either an absorption or emission feature relative to the Cosmic Microwave Background (CMB), depending on the relative temperatures of the intergalactic medium (IGM) and the CMB. Its precise shape reflects the properties of the first generations of stars, galaxies, and black holes, offering a direct probe into early astrophysical phenomena. However, detecting such a faint signal, which has an expected depth of only $\lesssim \SI{200}{mK}$ \cite{dhandha2025narrowing}, poses many new scientific and engineering challenges. The signal is buried deep beneath foreground emissions from our galaxy and extragalactic sources, which are up to five orders of magnitude stronger, as depicted in Fig.~\ref{fig:21cm_signal}. This dynamic range disparity demands high-precision instruments and rigorous calibration techniques, in order to isolate the cosmological component.

\begin{figure}[!htb]
\centering
\begin{subfigure}{.44\columnwidth}
  \centering
  \includegraphics[width=.98\linewidth]{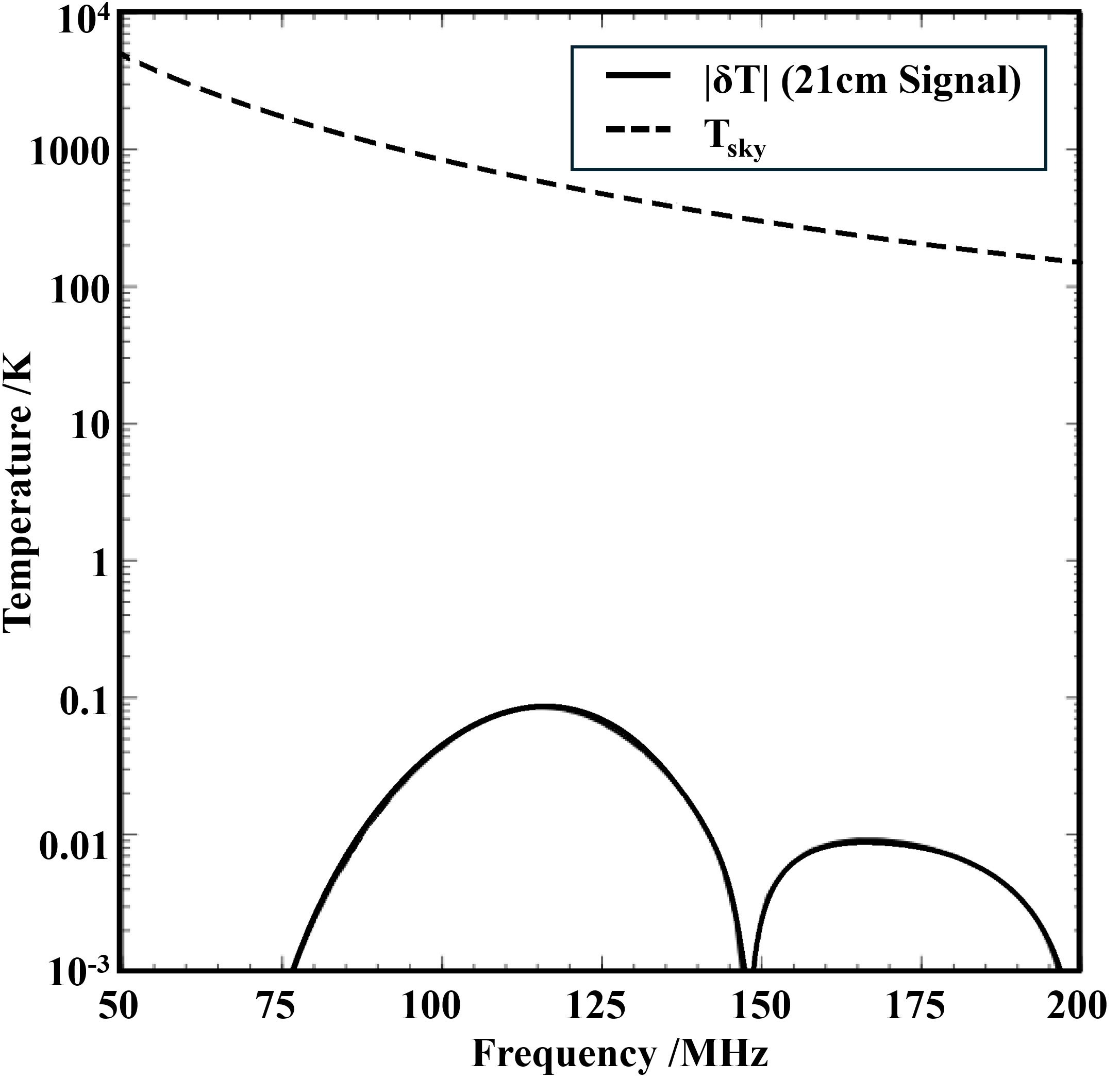}
  \caption{}
  \label{fig:21cm_signal}
\end{subfigure}
\begin{subfigure}{.55\columnwidth}
  \centering
  \includegraphics[width=.98\linewidth]{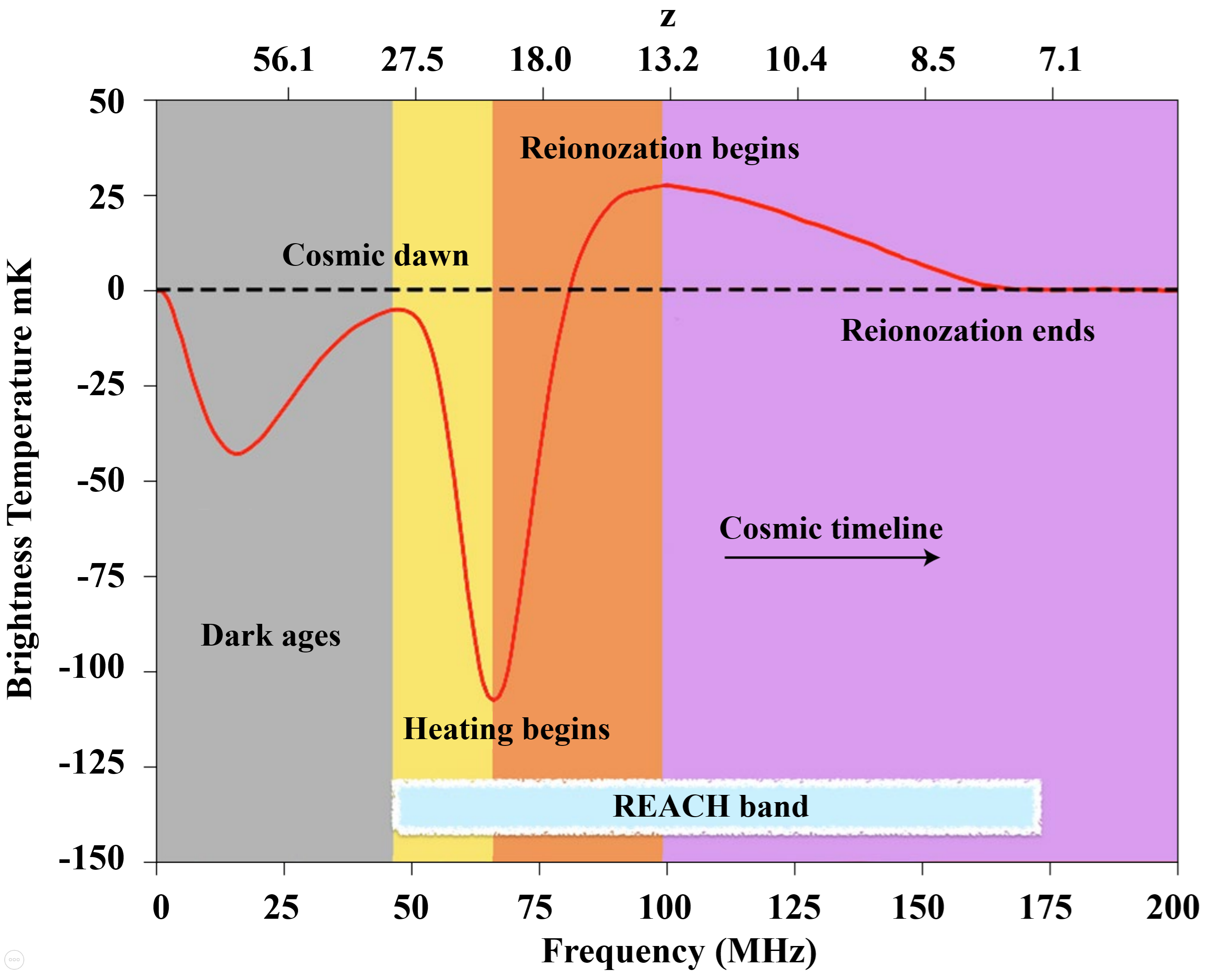}
  \caption{}
  \label{fig:21cm_line}
\end{subfigure}
\caption{Models of the sky and signal temperatures, showing (\subref{fig:21cm_signal}) the magnitude of the expected sky temperature, modeled as a -2.5-power law, against the absolute magnitude of an example global 21 cm signal (adapted from \citet{Cumner2022CMB}), and (\subref{fig:21cm_line}) a typical model of the global 21 cm signal, with (from left to right): collisional coupling (grey), onset of Ly-$\alpha$ coupling (yellow), onset of X-ray heating (orange) and photoionization (purple) (adapted from \citet{de2022reach}).}
\label{fig:21cm_signal}
\end{figure}

To detect the ``global'' 21 cm signal, many experiments, such as REACH \cite{de2022reach}, EDGES \cite{bowman2018absorption} and others \cite{SARAS, philip2019probing, sokolowski2015bighorns, voytek2014probing, price2018design, burns2019dark, monsalve2024mapper} employ radiometers composed of one or more single-element (non-interferometric) antennas designed to measure the sky-averaged radio spectrum. These instruments operate at frequencies corresponding to high redshifts ($z > 7.5$), typically in the 30–200 MHz range. Even though the instrumental concept is straightforward, extracting this global 21 cm signal requires an in-depth understanding of the system’s electromagnetic and thermal behavior. Small instrumental effects, such as impedance mismatches, cable losses, and temperature-dependent gain variations, can introduce spectral features that mimic or obscure the target signal. Calibration strategies must therefore carefully understand, model and account for any of these effects, while avoiding any possibility of fitting out the signal. As a case study, this paper focuses on the Radio Experiment for the Analysis of Cosmic Hydrogen (REACH): a global 21 cm experiment operating in the frequency band shown in Fig.~\ref{fig:21cm_line} in the Karoo Radio Reserve in South Africa. REACH aims to verify or refute the controversial 21 cm signal detection reported by the EDGES experiment \cite{bowman2018absorption}, which claimed evidence of a strong absorption feature that has significant implications for both astrophysics and fundamental physics. To ensure the reliability of its findings, REACH employs a calibration strategy with strong focus on hardware design, environmental monitoring, and system modeling. However, the in situ characterization of some of the corrections required by the project, such as those accounting for calibration source temperature differences caused by losses in the signal chains, is not possible using existing methods.

In this work, we present a novel methodology used to model the REACH calibration sources and predict otherwise immeasurable systematics. Traditionally, such systematics are accounted for using previously taken laboratory measurements, which may not accurately reflect the instrument's performance in its operational environment. A more robust approach is to employ in situ measurements. By measuring S-parameters, temperatures, and other system metrics, and then integrating these readings into the calibration pipelines, it is possible to account for drift in the system over time. The proposed framework integrates traditional circuit modeling with a combination of frequentist optimization and Bayesian inference techniques to fit for REACH's calibration source S-parameters. This approach allows for efficient and iterative modeling, robust parameter estimation, increased system understanding, and detailed uncertainty quantification of predictions. Although we focus on the case study of temperature corrections for the calibration sources, the methodology lays the groundwork for other applications in dynamic radiometer calibration and system monitoring. The result is a general set of techniques, that can ultimately be used to enhance the accuracy and robustness of future experiments.

The structure of this paper is as follows. Section \ref{sec:front_end} provides an overview of the REACH radiometer and the project's strategies, as well as the motivation for in situ calibration. Section \ref{sec:methodology} describes the methodology developed in this work, which includes a description of traditional circuit modeling, the frequentist and Bayesian techniques, and a general pipeline that summarizes the overall approach. Section \ref{sec:source_modeling} and \ref{sec:results} then applies this framework to the REACH calibration sources for the motivating example of the temperature corrections, describing the specific models in detail and analyzing results obtained from a combination of simulated and measured data. Finally, the paper is concluded in Section \ref{sec:conclusion}, along with suggestions for future work.
\section{REACH Radiometer}\label{sec:front_end}
The REACH telescope is a wideband radiometer designed to detect the spatially and temporally averaged global 21 cm signal. Operating across a frequency range of 50–170 MHz (corresponding to a redshift range of approximately 7.5 to 28), it aims to probe the signals from both the Cosmic Dawn and the Epoch of Reionization. REACH is a fully automated, standalone radiometric system, engineered for long-term, unattended deployment in remote and radio-quiet locations such as the Karoo, South Africa.

\subsection{Front-end Receiver}
\begin{figure}
\centering
\begin{subfigure}{1.0\columnwidth}
\centering
\includegraphics[width=0.85\columnwidth]{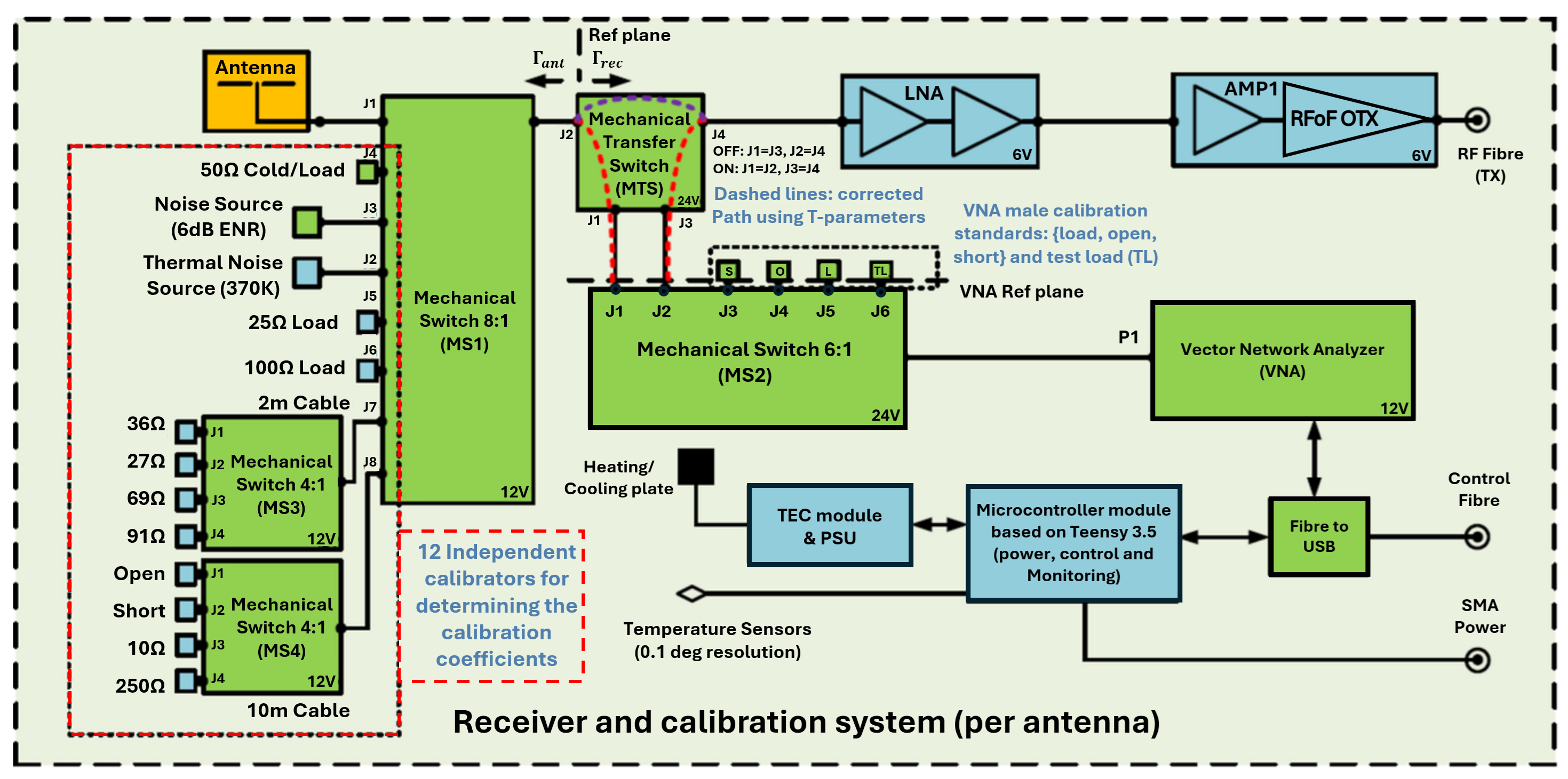}
\caption{}
\label{fig:reach_frontend}
\end{subfigure}\
\begin{subfigure}{1.0\columnwidth}
\centering
\includegraphics[width=0.85\columnwidth]{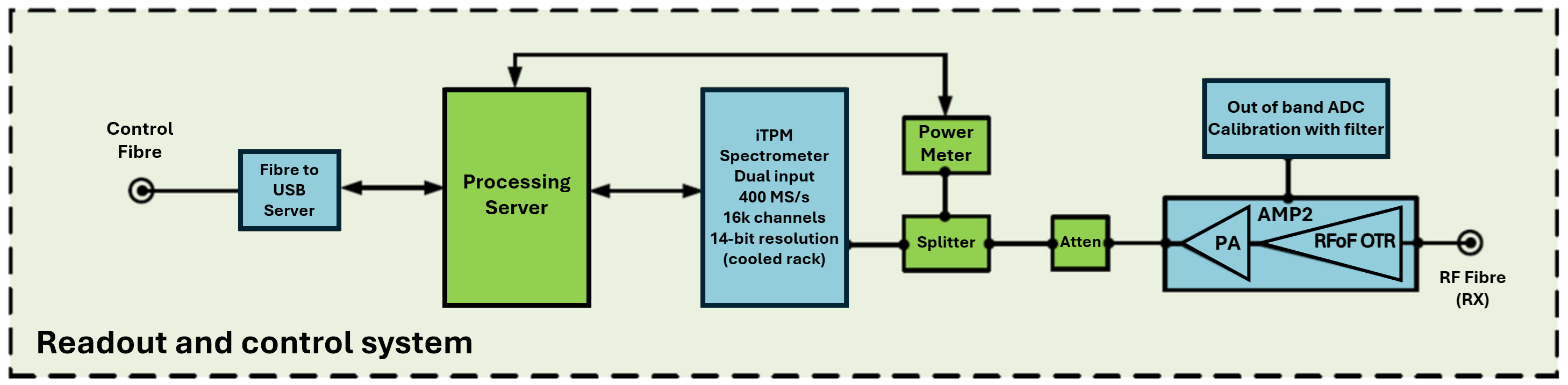}
\caption{}
\label{fig:reach_backend}
\end{subfigure}
\caption{An overview of the REACH radiometer, showing (\subref{fig:reach_frontend}) the front-end receiver with 12 independent calibration sources and the antenna connected to the receiver via mechanical switches, and (\subref{fig:reach_backend}) the back-end receiver control system and RF-over-fibre link (adapted from \citet{roque2025receiver}).}
\label{fig:reach_receiver}
\end{figure}

A simplified block diagram of the REACH front-end receiver is presented in Fig.~\ref{fig:reach_frontend}. The system incorporates twelve independent calibration sources with various reflection coefficients ($S_{11}$ parameters) in order to provide redundancy to the calibration algorithms. In this work, these are referred to as ``sources'', where each source is connected to the measurement systems (for power and S-parameter measurements) through a chain of connectors, coaxial cables, and switches. The loads consist of either high-quality, off-the-shelf \SI{50}{\ohm} matched terminations (used for the hot and cold sources), standard open and short terminations, or custom-made loads where the connector is terminated with precision resistors. One of these loads, referred to as the ``hot load'', is heated above ambient temperature by means of a heater affixed to the high-quality matched termination. The receiver also incorporates a calibrated noise source and a low-noise amplifier (LNA), whose input can be switched between the various sources and the antenna.


In order to enable in situ measurements, the front-end receiver includes a portable vector network analyzer (VNA) for measuring the S-parameters of the system. Additionally, a thermocouple module continuously monitors the internal temperature of the receiver, the cables, and each source, to ensure stable operating conditions. The receiver's temperature is actively regulated through a Peltier module, which is controlled by the onboard microcontroller. This microcontroller manages the operation of the RF switch network, automating calibration sequences and ensuring that the thermal environment is maintained within optimal limits.

The various sources are connected to the network analyzer through high-quality mechanical RF switches as well as a combination of semi-rigid and flexible coaxial cables. While the switches have signal path insertion losses typically as low as \SI{0.01}{\decibel}, the losses in the cables are non-negligible due to their physical length. Mismatches among all system components influence the noise power ultimately received by the spectrum analyzer and must be taken into account. A detailed technical description of the REACH receiver is provided in \citet{roque2025receiver}.

\subsection{Data Analysis Strategy}
REACH distinguishes itself from previous global 21 cm experiments through its architecture, with a strategy designed to decouple the effects of the instrumental response, astrophysical foregrounds, and the underlying cosmological signal. Although treating these three components as orthogonal subspaces simplifies the analysis \cite{de2022reach, Dominic_antenna_design}, perfect orthogonality is not achievable in practice, and degeneracies between them inevitably arise. To address this, REACH has employed a Bayesian calibration and data analysis framework \cite{roque2021bayesian, Dominic_chromaticity_correction, Michael_map_errors} that enables statistical characterization and mitigation of correlations between these components. 

This approach is further supported by adopting physically motivated models for both the instrument and the foreground. These models are constrained by direct physical measurements acquired during field operations. For example, the reflection coefficient of each calibration source is measured and incorporated directly into the calibration pipeline. This integration of dynamic measurements represents a relatively new approach in radio astronomy, in which the performance of the system is continuously monitored and adjusted on the basis of real-time data. In addition, by using empirically grounded system models, the project aims to enhance both the robustness and credibility of the final signal extraction and inference.

\subsection{Calibration Strategy}\label{sec:calibration_strategy}
\begin{figure}[!htb]
    \centering
    \includegraphics[width=0.85\columnwidth]{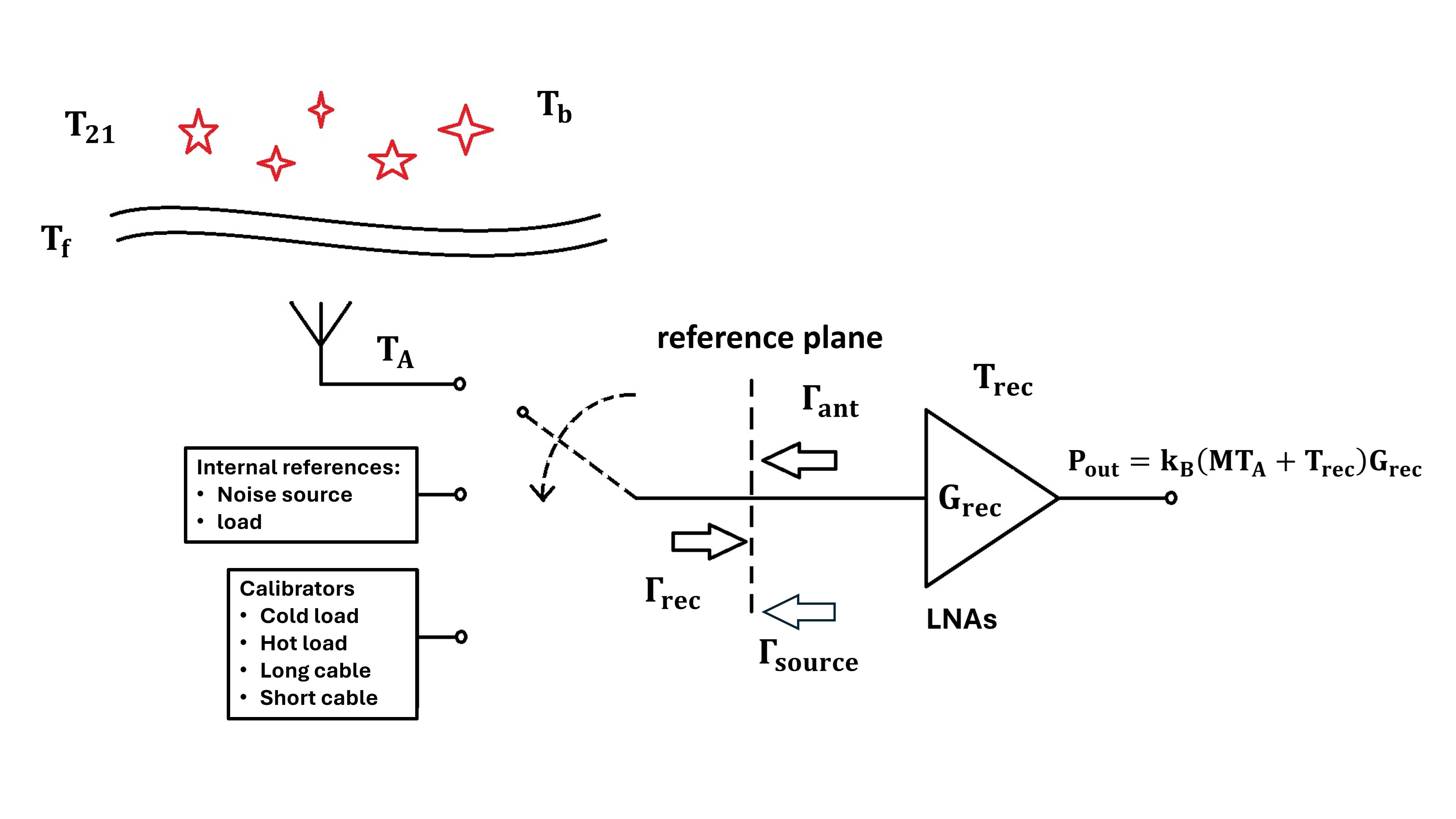}
    \caption{An illustration of the switching procedure between the antenna and the calibration sources, which connect to the input of the radiometric receiver. $\Gamma_\mathrm{rec}$ represents the reflection coefficient observed looking into the receiver, while $\Gamma_\mathrm{ant}$ and $\Gamma_\mathrm{source}$ are the reflection coefficients looking back into the antenna and calibration sources.  $T_\mathrm{A}$ represents the antenna temperature, comprised of the global 21 cm signal, $T_\mathrm{21}$, as well as foreground and background components, $T_\mathrm{f}$ and $T_\mathrm{b}$. When the antenna is connected, the output power, $P_\mathrm{out}$, consists of a sum of the mismatched antenna and internal receiver temperatures, $M T_\mathrm{A}$ and $T_\mathrm{rec}$, multiplied by Boltzmann's constant,  $k_\mathrm{B}$, and the receiver's gain, $G_\mathrm{rec}$.}
    \label{fig:dicke_switching}
\end{figure}

In the REACH system, a calibration strategy has been adopted that dynamically fits for the receiver systematics. Fig.~\ref{fig:dicke_switching} illustrates a conceptual block diagram of the signal pathways and interactions present. The antenna temperature, denoted as $T_\mathrm{A}$, is the beam-weighted sum of the astrophysical foregrounds, $T_\mathrm{f}$, the cosmological 21 cm signal, $T_{21}$, and other background sources, $T_\mathrm{b}$. In the context of the calibration pipelines, the goal is to provide a reliable measurement of $T_\mathrm{A}$, such that further data analysis can be used in attempt to detect the extremely faint $T_{21}$.

Several non-idealities exist in the signal chains. Firstly, fluctuations in the amplifier gain, $G_\mathrm{rec}$, lead to inconsistent measurements. Secondly, all lossy and active components contribute additional thermal noise that needs to be quantified. Finally, several impedance mismatches are present at the reference plane defined at the LNA's input, which introduce additional complexity. Referring to the diagram, $\Gamma_\mathrm{ant}$ is known to vary greatly with frequency, while $\Gamma_\mathrm{source}$ is also frequency-dependent and changes depending on the source connected. These reflections, quantified in a mismatch factor, $M$, lead to signal distortions, standing waves, and additional contributions from thermal noise which must be characterized. In order to address some of these issues, REACH utilizes a variant of a calibration approach based on Dicke switching.

\subsubsection{Dicke Switching}
Dicke switching was first suggested as a means of compensating for any gain variations in the receiver \cite{dicke1946measurement}. In this approach, the receiver input is continuously switched between the sky-facing antenna and internal calibration sources during an observation. By using only one reference load, and a switching frequency much higher than the typical timescale of the gain fluctuations, the effective gain becomes flat over the timescale of a complete switching cycle, compensating for any long-term drift in the gain. Further, if a second reference load with known noise characteristics is introduced, then an absolute calibration of both the LNA gain and its internal noise temperature can be achieved. However, traditional Dicke switching assumes that the antenna and calibration sources are well-matched to the receiver. For modern, wideband, high-precision experiments such as REACH, this is typically not the case, and more advanced methods, such as noise wave parameters, are needed. 

\subsubsection{Noise Wave Parameters}
One method proposed to calibrate for source-dependent reflections is that of noise wave parameters, introduced in \citet{meys1978wave} and adapted to absolute radiometric calibration in \citet{rogers2012absolute}. Although newer strategies exist (for example, variations that employ machine learning to fit for so-called \textit{noise parameter surfaces} \cite{leeney2025radiometer}), this work focuses on the original formalism that was taken into account in the design of the REACH receiver.

In the noise wave parameterization, the final power measured by the spectrometer, $P_\mathrm{source}$, is written as a function of the temperature of the calibration source, $T_\mathrm{source}$, the reflection coefficients, $\Gamma_\mathrm{source}$ and $\Gamma_\mathrm{rec}$, and a set of ``noise wave'' parameters, resulting in
\begin{align}\label{eqn:noise_wave_power}
    T_\mathrm{NS} \left(\frac{P_{\mathrm{source}} - P_L}{P_{\mathrm{NS}} - P_\mathrm{L}}\right) + T_L &= T_{\mathrm{source}}\left[ \frac{1 - |\Gamma_\mathrm{source}|^2}{|1-\Gamma_\mathrm{source}\Gamma_\mathrm{rec}|^2} \right] \nonumber \\
    &+ T_\mathrm{unc} \left[ \frac{|\Gamma_\mathrm{source}|^2}{|1-\Gamma_\mathrm{source}\Gamma_\mathrm{rec}|^2} \right] \nonumber \\
    &+ T_\mathrm{cos} \left[ \frac{\mathrm{Re}\left( \frac{\Gamma_\mathrm{source}}{1 - \Gamma_\mathrm{source}\Gamma_\mathrm{rec}} \right)}{\sqrt{1 - |\Gamma_\mathrm{rec}|^2}} \right] \nonumber \\
    &+ T_\mathrm{sin} \left[ \frac{\mathrm{Im}\left( \frac{\Gamma_\mathrm{source}}{1 - \Gamma_\mathrm{source}\Gamma_\mathrm{rec}} \right)}{\sqrt{1 - |\Gamma_\mathrm{rec}|^2}} \right],
\end{align}
where $T_\mathrm{unc}$, $T_\mathrm{cos}$ and $T_\mathrm{sin}$ are the noise wave parameters, and $P_\mathrm{L}$, $P_\mathrm{NS}$, $T_\mathrm{L}$ and $T_\mathrm{NS}$ are the powers and temperatures associated with a reference load and noise source respectively. In general, all values in (\ref{eqn:noise_wave_power}) are frequency-dependent. Ultimately, these parameters quantify the effect of the noise reflections due to the LNA's non-idealities. Since this equation is linear in the unknowns, it can be solved for given sufficient measurements of the source reflections. Note that the measurements required by this equation are traditionally taken in the laboratory, but are made in situ in the REACH case. Finally, with the unknowns calculated, the same equation can then be used to infer the antenna temperature, $T_\textnormal{A}$, given a measurement of its power, $P_\textnormal{A}$.

Although the above formalism allows for accurate characterization of the LNA, it relies on several assumptions and corrections. Firstly, all reflection coefficients measurements are assumed to be with respect to the main reference plane (indicated as ``Ref plane'' in Fig.~\ref{fig:reach_frontend}). In reality, this is not the case, and S-parameter corrections (embedding/de-embedding) are required. Secondly, losses in the signal chains (mostly caused by the cables) affect the equivalent source temperatures, $T_\mathrm{source}$, required for (\ref{eqn:noise_wave_power}), and therefore a corresponding temperature correction is required before calibration.

\subsection{Temperature Corrections}\label{sec:temperature_corrections}
\begin{figure}[!htb]
    \centering
    \includegraphics[width=0.7\linewidth]{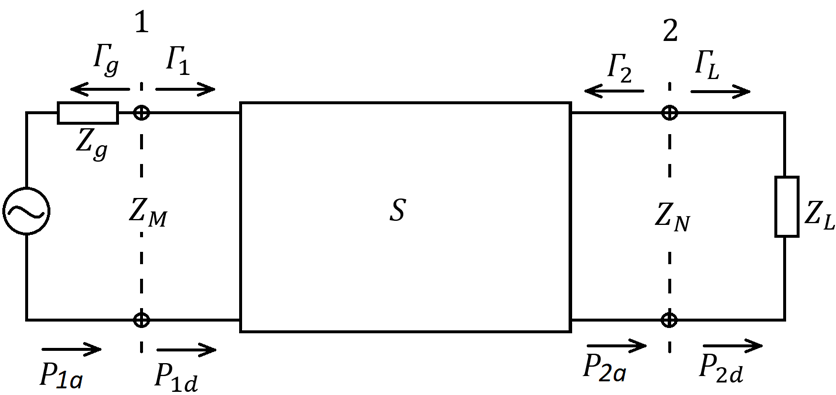}
    \caption{A circuit diagram depicting the various reflection coefficients and mismatch impedances for an arbitrary noise generator connected via a two-port network to a load (adapted from \citet{miller1967noise}). $P_\mathrm{1a}$/$P_\mathrm{1d}$ and $P_\mathrm{2a}$/$P_\mathrm{2d}$ are the available/delivered powers on the left/right of reference planes 1 and 2 respectively.}
    \label{fig:gain_network}
\end{figure}

The motivating case study for this work is the temperature correction required for the REACH sources. To derive the required correction, a general circuit diagram for a noise generator with internal impedance $Z_\mathrm{g}$ connected via a two-port network to an arbitrary load, $Z_\mathrm{L}$, is shown in Fig.~\ref{fig:gain_network}. The two-port network is represented by its scattering matrix, $[S]$. The associated reflection coefficients of the generator and absorbing load, $\Gamma_\mathrm{g}$ and $\Gamma_\mathrm{L}$, are also shown, as well as those looking into the terminated two-port network, $\Gamma_1$ and $\Gamma_2$, at reference planes 1 and 2 respectively. The characteristic line impedances, $Z_\mathrm{M}$ and $Z_\mathrm{N}$, are also defined at these planes which, in general, need not be the same.

As detailed in \citet{miller1967noise}, the equivalent noise temperature at reference plane 2 (the output of the two-port) is a function of the input noise temperature of the generator, as well as the impedance mismatches and losses of the system. The mismatch factors at planes 1 and 2 are
\begin{align}
    M &= \frac{(1 - |\Gamma_\mathrm{g}|^2)(1 - |\Gamma_1|^2)}{|1-\Gamma_\mathrm{g} \Gamma_1|^2},    & 
    N &= \frac{(1 - |\Gamma_\mathrm{L}|^2)(1 - |\Gamma_2|^2)}{|1-\Gamma_\mathrm{L} \Gamma_2|^2}.
\end{align}
The power gain, relating the \textit{delivered} powers, is
\begin{equation}
    G_\mathrm{d} = \frac{P_\mathrm{2d}}{P_\mathrm{1d}} = \frac{(Z_\mathrm{M}/Z_\mathrm{N}) |S_{21}|^2(1-|\Gamma_\mathrm{L}|^2)}{(1-|\Gamma_1|^2)|1-S_{22}\Gamma_\mathrm{L}|^2}.
\end{equation}
The available gain, relating the \textit{available} powers, is
\begin{equation}
    G_\mathrm{a} = \frac{P_\mathrm{2a}}{P_\mathrm{1a}} = \frac{M}{N} G_\mathrm{d} = \frac{(Z_\mathrm{M}/Z_\mathrm{N})(1 - |\Gamma_\mathrm{g}|^2)|S_{21}|^2}{(1 - |\Gamma_2|^2)|1-S_{11}\Gamma_\mathrm{g}|^2}.
\end{equation}
If the resultant network is matched and lossless, then $G_\mathrm{d} = G_\mathrm{a} = 1$. However, in the general lossy case, these gains can take on any value between $0$ and $1$, and are typically functions of frequency.

\begin{figure}[!htb]
    \centering
    \includegraphics[width=0.8\linewidth]{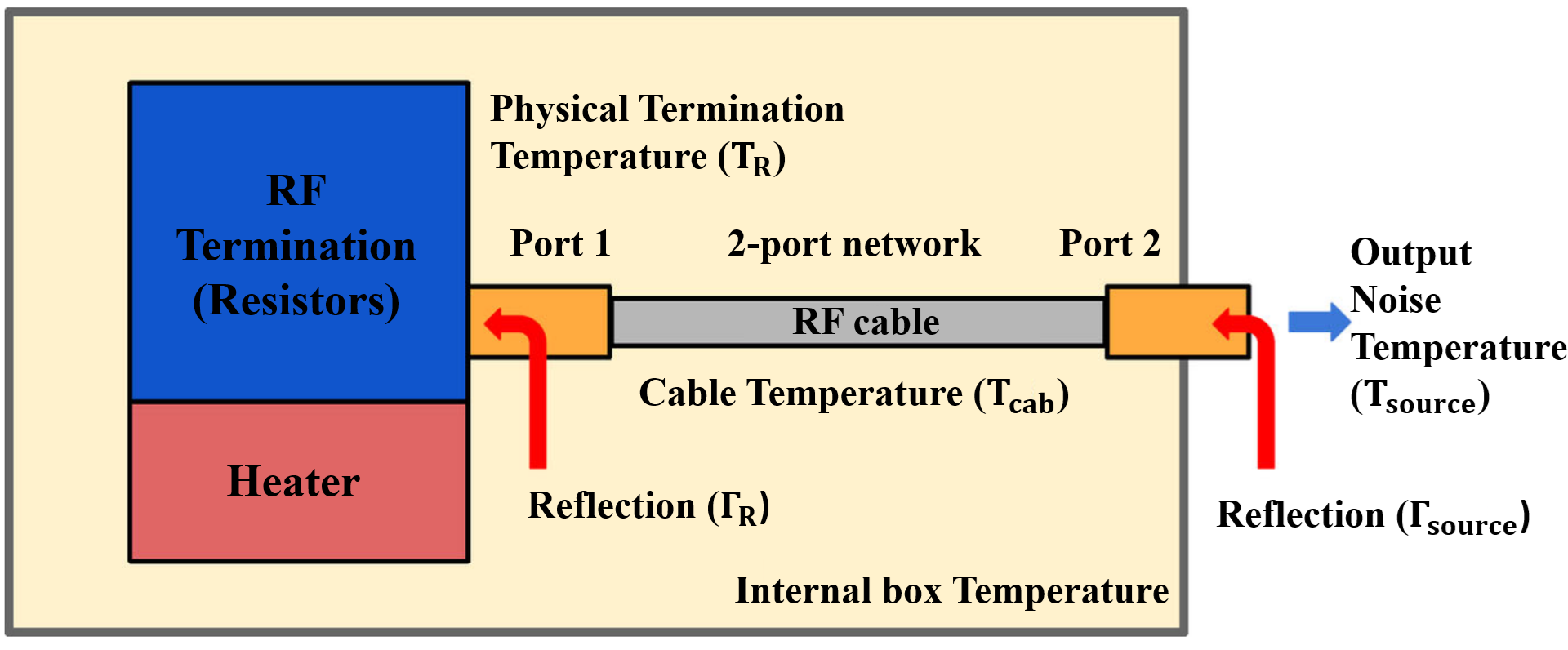}
    \caption{An illustration of a typical source (heated load) connected to the receiver via an RF cable, with load reflection $\Gamma_R$, load and cable temperatures $T_R$ and $T_\mathrm{cab}$, source reflection $\Gamma_\mathrm{source}$, and effective temperature $T_\mathrm{source}$ (adapted from \citet{roque2025receiver}). Note that $T_\mathrm{R}$, $T_\mathrm{cab}$ and $T_\mathrm{source}$ are not necessarily equal.}
    \label{fig:heated_load}
\end{figure}

The available power of a noisy source, $P_\mathrm{1a}$, as a function of its equivalent noise temperature, $T_\mathrm{g}$, is
\begin{equation}
    P_\mathrm{1a} = k_\mathrm{B} T_\mathrm{g} B
\end{equation}
where $k_\mathrm{B}$ is Boltzmann's constant and $B$ is the system bandwidth. Assuming the two-port network is at a constant ambient temperature, $T_\mathrm{0}$, then the effective or available noise temperature at reference plane 2, $T_\mathrm{eff}$, is ultimately a weighted sum of the noise from the generator, and the noise in the two-port itself,
\begin{equation}
    T_\mathrm{eff} = G_\mathrm{a} T_\mathrm{g} + (1 - G_\mathrm{a}) T_\mathrm{0}.
\end{equation}
In the context of the REACH radiometer, Fig.~\ref{fig:gain_network} is representative of any one of the 12 calibrator chains. Specifically, the generator reflection coefficient, $\Gamma_\mathrm{g}$, and noise temperature, $T_\mathrm{g}$, relate to each calibration load's reflection, $\Gamma_\mathrm{R}$, and physical temperature, $T_\mathrm{R}$; $[S]$ represents the source's signal chain; and $Z_\mathrm{L}$ is associated with the impedance of the LNA itself. This arrangement is illustrated for the hot source in Fig.~\ref{fig:heated_load}.

It is now assumed that all S-parameters are with respect to a consistent characteristic impedance (\textit{e.g.} $\SI{50}{\ohm}$) such that $Z_\mathrm{M} = Z_\mathrm{N}$. It is also assumed that the entire signal chain, $[S]$, is at the same physical temperature as the cable, $T_\mathrm{cab}$, in order to simplify the analysis. However, if the connectors and switches are at different temperatures, the formalism in \citet{mukaihata1968applications} can be used. Given the reflection coefficients, physical temperatures, and signal chain's $S_{11}$ and $S_{21}$, the gain and effective source temperature are
\begin{align}\label{eqn:temperature_corrections}
    G_\mathrm{a} &= \frac{(1 - |\Gamma_\mathrm{R}|)^2|S_{21}|^2}{(1-|\Gamma_\mathrm{source}|^2)|1-S_{11}\Gamma_\mathrm{R}|}, &
    T_\mathrm{source} = G_\mathrm{a} T_\mathrm{R} + (1 - G_\mathrm{a}) T_\mathrm{cab}.
\end{align}

Note that, although the final noise power delivered to the load is actually $N T_\mathrm{source}$, it is only $T_\mathrm{source}$ itself that must be substituted into (\ref{eqn:noise_wave_power}), since the mismatch factor between the full source and the receiver is implicit in the derivation of the noise wave calibration.

\subsection{Motivation for In Situ Calibration}
A central design principle of the REACH receiver is to enable in situ calibration through an integrated vector network analyzer (VNA), distributed temperature probes, and a spectrometer, providing real-time estimates of the parameters in (\ref{eqn:noise_wave_power}). However, for the temperature correction of (\ref{eqn:temperature_corrections}), not all parameters can be measured in situ. While real-time measurements of $T_\mathrm{R}$, $T_\mathrm{cab}$, and $\Gamma_\mathrm{source}$ are readily available, the correction still depends on laboratory measurements of the signal chain $S_{11}$ and $S_{21}$, and load $\Gamma_R$. These cannot be obtained without dismantling the system, which would undermine the principle of true in situ calibration. The work presented here overcomes this limitation, enabling real-time predictions of these immeasurable quantities and therefore dynamic corrections of systematics. With this contribution, the three key advantages of in situ calibration over conventional methods are summarized:
\begin{arabiclist}
    \item \textit{Accurate values at observation time}. By measuring calibration sources during observations, REACH ensures that values such as reflection coefficients and temperatures, as required by (\ref{eqn:noise_wave_power}), are directly relevant to the prevailing environment and system conditions.
    \item \textit{System monitoring}. Direct analysis of the measurements, as well as of model fits, can be used to monitor any long-term drift or degradation of the system.
    \item \textit{Real-time corrections}. By employing physical models, immeasurable parameters can be predicted, and corrections that are usually computed using previous laboratory data can instead be informed by real-time measurements. \label{itm:corrections}
\end{arabiclist}
The scope of this work is limited to the final use case of \ref{itm:corrections}, specifically for temperature corrections, due to an analysis that revealed the sensitivity of the source temperatures to perturbations in the system parameters (demonstrated in Section \ref{sec:source_modeling}). However, the techniques can easily be applied to other use-cases, such as dynamic characterization of the instrument's antenna (using \textit{e.g.} the REACH balun circuit model developed in \citet{allen2025balun}); in-depth monitoring and fault detection; and the S-parameter corrections mentioned previously.
\section{Methodology}\label{sec:methodology}
This section presents the methodology in a general manner, before specifically applying it to the REACH sources in Sections \ref{sec:source_modeling} and \ref{sec:results}. The circuit modeling approach is explained, as well as a commonly-used foundational transmission line model. The frequentist and Bayesian techniques are also described, to be used for model fitting and the in-depth analysis of results. The section is concluded with the description of a unified pipeline, which integrates the strengths of the above approaches in a way that practically enables both efficient modeling and robust statistical inference.

\subsection{Circuit Modeling}
Electromagnetic circuit modeling has long been an important technique in RF and microwave engineering. Microwave circuit models are physically-based ``macromodels'' that represent a complex electromagnetic system using a simplified, equivalent circuit \cite{pozar2012microwave}. These circuits typically include a combination of ``lumped'' elements, such as resistors, capacitors and inductors, and ``distributed'' elements, such as transmission lines. Such elements are either used to directly model their physical counterpart, for example resistors or cables, or to capture any ``parasitic'' electromagnetic fields (either capacitive or inductive in nature). By combining these elements in a physically motivated manner, arbitrarily complex circuit models can be envisioned that capture the core properties of an RF component.

Typically, such models are represented using equations written in terms of the S- or ABCD- matrices of the individual components. While lumped elements have simple mathematical forms, models for distributed components, such as coaxial cables or printed circuit board (PCB) transmission lines (\textit{e.g.} microstrip or stripline), have complex equations depending on the specific geometry. In the case of 21 cm experiments, however, such lines are usually approximately transverse electromagnetic (TEM) in nature, in which case a foundational RLGC transmission line model can be utilized.

\subsubsection{RLGC Model}
\begin{figure}
    \centering
    \includegraphics[width=0.5\linewidth]{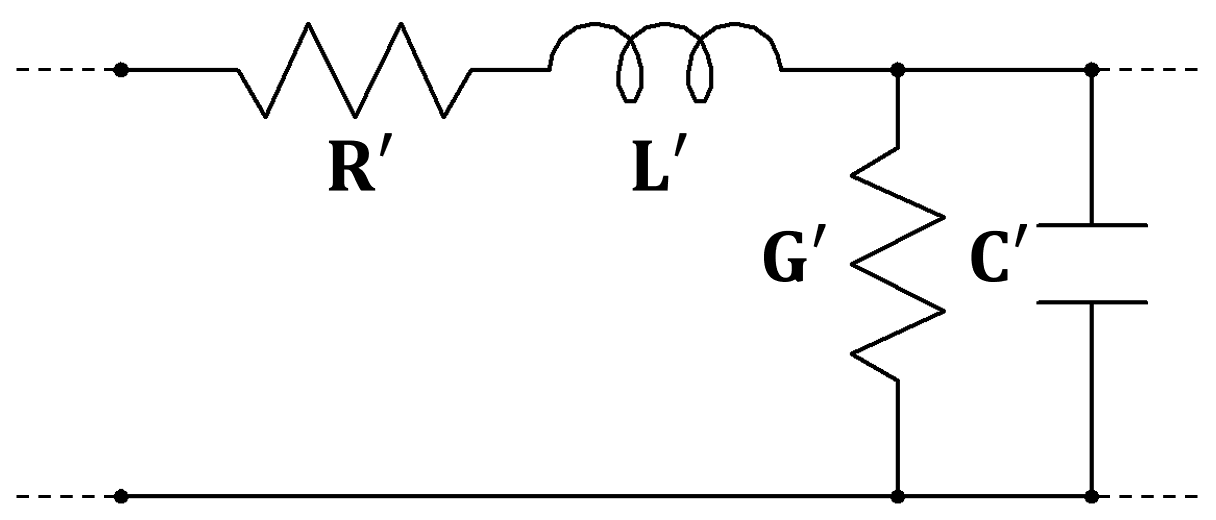}
    \caption{One section of an RLGC transmission line.}
    \label{fig:rlgc}
\end{figure}

The RLGC model, described in detail in \citet{pozar2012microwave}, treats a transmission line as a cascade of infinitely many, infinitesimally small series and shunt lumped-element sections, consisting of resistance (R), inductance (L), conductance (G), and capacitance (C). The circuit layout for a single RLGC section is shown in Fig.~\ref{fig:rlgc}, with the line’s ``per-unit'' parameters $R'$, $L'$, $G'$ and $C'$ indicated. The model relies on two primary assumptions, namely that (i) only the TEM mode is present, and (ii) the transmission line is uniform along its length. In the context of global 21 cm experiments, the first assumption is valid due to the relatively low operating frequency. However, the second assumption can be violated by deformations or inhomogeneities in the specific transmission line. If this is found to be significant, augmentation of the model with more data-driven approaches may be necessary, though this is not addressed in this work.

As demonstrated in later sections, this RLGC model is capable of exactly representing any uniform TEM line given its per-unit parameters. This contrasts with models that use further approximations, for example those employing the common high-frequency, low-loss simplification, which can produce significant errors \cite{djordjevic2003note}. Such errors were observed to be non-negligible in the case of the REACH balun circuit model, and so similar approximations should be carefully criticized.

To compute an RLGC line's two-port parameters, a general TEM line is first considered. For a given characteristic impedance, $Z_c$, propagation constant, $\gamma$, and physical length, $\ell$, the ABCD matrix is
\begin{equation}\label{eqn:abcd_tem}
    \mathbf{A}_{\textnormal{TEM}} =
    \begin{bmatrix}
        \cosh\gamma \ell                      &       Z_c \sinh{\gamma \ell} \\
        \frac{1}{Z_c}\sinh{\gamma \ell}         &       \cosh{\gamma \ell} \\
    \end{bmatrix},
\end{equation}
where all values (except the length) are, in general, complex. Next, it is assumed that the line is operating at an angular frequency $\omega = 2 \pi f$. Kirchoff's laws can then be used to derive differential equations for the voltage and current of the line, resulting in the final formulation
\begin{align}
    Z_c (\omega) &= \sqrt{\frac{R'(\omega) + j \omega L'(\omega)}{G'(\omega) + j \omega C'(\omega)}},
    \label{eqn:Z_c}  \\ 
    \gamma (\omega) &= \sqrt{(R'(\omega) + j \omega L'(\omega)) \cdot (G'(\omega) + j \omega C'(\omega))},
    \label{eqn:gamma} 
\end{align}
where $j = \sqrt{-1}$. Therefore, given the length and per-unit RLGC parameters for an arbitrary TEM line as a function of frequency, its ABCD parameters can be calculated.

\subsubsection{Implementation}
While commercial, GUI-driven circuit modeling tools exist, they often lack the programmatic interface needed to easily define custom models or cost functions, and are unable to integrate with modern optimizers or Bayesian samplers. A custom, open-source framework, ParamRF \cite{allen2025paramrf}, has therefore been developed, which caters for efficient, declarative circuit modeling and fitting in Python.

\subsection{Frequentist Optimization}\label{sec:frequentist_optimization}
To initially develop and verify various circuit models against a large number of datasets, it is useful to have an efficient frequentist optimization strategy. Ideally, this should be both robust and flexible, utilizing popular minimization routines and cost functions. In practice, however, optimal convergence is made difficult due to several reasons. These include the need to fit multiple measurements simultaneously; the large number of parameters involved; and degeneracies and correlations in the parameter space. Simultaneous measurements pose the biggest challenge, primarily due to varying scales in the data. For example, when fitting two-port measurements of a well-matched RF component, residuals for both the reflection and transmission coefficients should be minimized. However, typically $S_{11}\approx S_{22}\approx 0$, whereas $S_{12} \approx S_{21}\approx 1$. Additionally, in the REACH case, various components are shared between sources due to switching (for example, the $\SI{10}{m}$ cable between the four sources on the MS4 switch in Fig.~\ref{fig:reach_receiver}). Therefore, all 12 source reflection coefficients should be fit both simultaneously and in a ``joint'' manner \textit{i.e.} with model parameters appropriately shared. This hinders convergence further. To this end, a custom convolution-based cost function has been developed, which was found to be the most robust, outperforming more typical formulations in both accuracy and convergence.

\subsubsection{Traditional Cost Functions}
Let ${\tilde{\mathbf{s}}}_i$ and $\hat{\mathbf{s}}_i(\theta)$ denote the measured and modeled S-parameters respectively, with $i \in [1, n]$ and $n=12$ for the source reflection coefficient measurements in REACH. Here, the dimension of the vectors represents samples in frequency. Then, the complex and magnitude residual vectors can be written as
\begin{align}
    \mathbf{c}_i(\theta) &= \tilde{\mathbf{s}}_i - \hat{\mathbf{s}}_i(\theta), &
    \mathbf{m}_i(\theta) &= |\tilde{\mathbf{s}}_i| - |\hat{\mathbf{s}}_i(\theta)|,
\end{align}
where $\theta$ represents the model parameters, and $|\mathbf{x}|$ denotes the element-wise absolute value of $\mathbf{x}$. One approach often used is to minimize the sum of the $L_2$ norms of the complex or magnitude differences, resulting in the cost function definitions
\begin{align}\label{eqn:freq_loss}
    J_\mathrm{c}(\theta) &= \sum_{i=1}^n ||\mathbf{c}_i(\theta)||_2, &
    J_\mathrm{m}(\theta) &= \sum_{i=1}^n ||\mathbf{m}_i(\theta)||_2.
\end{align}
These cost formulations can easily result in the optimizer converging to local minima, with varying degrees of success depending on the starting point and the specific minimization algorithm. For example, in the case of $J_\mathrm{c}(\theta)$, since the magnitude of a complex difference is taken, any point, $\hat{\mathbf{s}}_i(\theta)$, that lies on a circular ring around the measured value, ${\tilde{\mathbf{s}}}_i$, will result in the same residual. Although this problem is mitigated as more frequency points are used, its exact effect will differ depending on the specific shape of the cost as a function of both frequency and the specific measurements.

One solution to address these issues is to use a weighted sum of the complex and magnitude costs. However, this introduces additional weight hyper-parameters into the routine. These have been found to be highly dependent on the starting point and the specific data being fit, making tuning such weights difficult.

\subsubsection{Convolutional Cost Function}
For the proposed convolutional cost function, instead of summing across the measurement axis, the individual L2 norms are stacked into complex and magnitude error vectors for each measurement as
\begin{align}
    \mathbf{c}(\theta) &= 
    \begin{bmatrix}
        ||\mathbf{c}_1(\theta)||_2 \\
        ||\mathbf{c}_2(\theta)||_2 \\
        \vdots \\
        ||\mathbf{c}_n(\theta)||_2 \\
    \end{bmatrix}, &
    \mathbf{m}(\theta) &= 
    \begin{bmatrix}
        ||\mathbf{m}_1(\theta)||_2 \\
        ||\mathbf{m}_2(\theta)||_2 \\
        \vdots \\
        ||\mathbf{m}_n(\theta)||_2 \\
    \end{bmatrix}.
\end{align}
The final cost is then the L2 norm of the convolution of these error vectors, written as
\begin{equation}\label{eqn:conv_cost}
    J_\textnormal{conv}(\theta) = ||\mathbf{c}(\theta) \ast \mathbf{m}(\theta)||_2.
\end{equation}
This composite metric is an automatically-weighted combination of magnitude and phase errors across all measurements, and has been found to effectively mitigates several convergence issues. Its effectiveness, however, does depend on the independence of the two vector spaces, which increases as more S-parameter measurements are used for the case of the REACH sources.

\subsubsection{Implementation}
The ParamRF framework caters for a number of minimization routines exhibiting varying performance depending on the models being fit. For high-dimensional, joint models such as the REACH sources, minimization routines found to perform well are typically those designed for constrained optimization, such as sequential quadratic programming algorithms or interior-point methods. In contrast, algorithms developed for unconstrained optimization, such as simplex methods or quasi-Newton approaches, tend to converge too slowly. However, when fitting simpler models such as cables, the converse is true, with constrained methods often converging too quickly to local minima. In practice, the SLSQP algorithm performs well in high dimensions, whereas the Nelder-Mead algorithm should be preferred for smaller problems.

\subsubsection{Limitations}
Although the frequentist approach provides good fits and fast convergence, it has several inherent limitations. For example, even though the convolutional cost function is able to combat the issues caused by more traditional cost functions, it had to be carefully designed. This will likely bias the resultant estimate of the parameters, which could influence model predictions. Also, the approach is still prone to local convergence, as well as degeneracies and biases within a given source chain. This not only can cause sub-optimal fits, but also incorrect parameters and therefore predictions. This is demonstrated in Section \ref{sec:results}. Further, frequentist techniques do not natively cater for the analysis of uncertainties in either the parameters or the predictions, which is of high priority in cosmological experiments.

\subsection{Bayesian Inference}\label{sec:bayesian_inference}
To combat the issues in the traditional frequentist approach, as well as provide further insight into (and control over) the models, this section details a Bayesian circuit fitting framework. The framework uses Bayesian inference to perform more robust parameter estimation, model comparison, and uncertainty analysis, as well as to update prior knowledge in a systematic manner.

Bayesian inference is a statistical method that reverses conditional probabilities through Bayes' thereom. Specifically, if $\theta$ is defined as the parameters for a model $\mathcal{M}$, and data $\mathcal{D}$ is collected, then
\begin{equation}\label{eqn:bayesianParameterEstimation}
    P(\mathbf{\theta} | \mathcal{D}, \mathcal{M}) = \frac{P(\mathbf{\theta} | \mathcal{M}) \cdot P(\mathcal{D} | \mathbf{\theta}, \mathcal{M})}{P(\mathcal{D} | \mathcal{M})}.
\end{equation}
The terms in the above equation have a common nomenclature,
\begin{equation}
    \textnormal{Posterior} = \frac{\textnormal{Prior} \cdot \textnormal{Likelihood}}{\textnormal{Evidence}}.
\end{equation}
This reads that the posterior probabilities of the parameters can be estimated by multiplying one's prior belief of those parameters with the likelihood that they align with the measured data (normalized by a term known as the evidence). ``Bayesian sampling'' algorithms can then be used to estimate the posterior, with some methods able to also compute the Bayesian evidence which can be used for model comparison. What follows is a description of the likelihood, priors and specific inference techniques used in this work.

\subsubsection{Likelihood}\label{sec:bayesian_likelihood}
For Bayesian inference, a likelihood function is required. In this context, this should combine some observational model with the circuit model. Here, ``observation'' refers to the network analyzer measurement. Often, VNA's make use of quadrature detection, where the in-phase (real) and quadrature (imaginary) components are measured separately. Using a matched load and a VNA, it is easy to verify that the noise in these components is independent and Gaussian, with a shared standard deviation, $\sigma$. Assuming this independence holds across measurements and frequency, the measured value, $\tilde{s}_{ij}$, can be related to the ``true'' value, $s_{ij}$, for S-parameter $i \in [1, n]$ and frequency $j \in [1, m]$ as
\begin{align}
    \tilde{s}_{ij} &= s_{ij} + \epsilon_{ij},   & \epsilon_{ij} \sim \mathcal{N}_\textnormal{re}(0, \sigma^2) + j\mathcal{N}_\textnormal{im}(0, \sigma^2),
\end{align}
where $\mathcal{N}$ represents a standard normal. $\tilde{s}_{ij}$ therefore follows a circularly symmetric, complex normal distribution centered on the true value, $s_{ij}$. The likelihood of observing a single measurement is
\begin{align}
    \mathcal{L}(\tilde{s}_{ij} |\sigma^2) = \frac{1}{\sqrt{2\pi\sigma^2}} \exp\left(-\frac{\left( \tilde{s}_{ij,\textnormal{re}} - s_{ij,\textnormal{re}}\right)^2 + \left( \tilde{s}_{ij,\textnormal{im}} - s_{ij,\textnormal{im}}\right)^2 }{2\sigma^2}\right). 
\end{align}
If a circuit model with $\hat{s}_{ij}(\theta)$ is assumed, then $s_{ij} \approx \hat{s}_{ij}(\theta)$, and the log-likelihood for $\theta$ and $\sigma$ is then
\begin{align}\label{eqn:logL_gaussian}   
    \log\mathcal{L}(\tilde{s}_{ij}|\theta, \sigma^2) = &-mn\log(2\pi \sigma^2) - \frac{1}{2\sigma^2} \sum_{i=1}^n \sum_{j=1}^m \left|\tilde{s}_{ij} -\hat{s}_{ij}(\theta) \right|^2.
\end{align}
Note that any non-negligible approximations in the model will violate the assumption that $s_{ij} \approx \hat{s}_{ij}(\theta)$, skewing the resultant standard deviation higher than the underlying measurement noise. Consequently, $\sigma$ also serves as a metric for model quality. A multi-variate Gaussian may therefore be used if multiple groups of $\sigma$ are desired (for example, to model the error of reflection and transmission coefficients separately). 

\subsubsection{Priors}\label{sec:bayesian_priors}
In general, any prior with a known inverse cumulative distribution (ICDF) may be used (as required by most sampling schemes), with the specific distribution chosen based on the experience of the engineer. However, a combination of Gaussian and uniform priors are likely sufficient, with Gaussian priors used for geometric parameters such as lengths, and uniform priors used for more heuristic parameters such as parasitic capacitances and inductances. Wider and less physical ``uninformative'' priors have also proven useful for circuit model evaluation and exploration, allowing the posterior to be predominantly data-driven.

The implementation of ``updated'' priors, however, is non-obvious (\textit{i.e.} using a previous set of posterior samples obtained from \textit{e.g.} laboratory data as new priors). While one approach is to use marginalized one-dimensional posteriors, this neglects dependencies between parameters. It has been found that exploiting such dependencies is important, for example to constrain one-port fits to the (degenerate) subspace allowed by previous two-port measurements. An approach found to work well is to use ``normalizing flows'', for example masked autoregressive flow (MAF) networks \cite{bevins2023marginal}. These are able to model multi-dimensional distributions directly, allowing full encapsulation of the two-port parameters of a specific RF component. By training the flow on posterior samples fit to two-port laboratory measurements, the resultant parameter space can be used as updated priors. Further, in the case when a given model is only to be used as a ``black box'' (\textit{i.e.} only S-parameter predictions are required), the flow can be trained on posteriors obtained from entirely uninformative priors, which combats any non-physicality in the model.

To account for parameter drift over time, posteriors can be ``broadened'' before being used as updated priors (though this is only necessary if the expected amount of drift has uncertainties greater than in the posterior). One method to achieve this is to perturb the posterior samples with Gaussian noise before training the flow network. The amount of perturbation added represents a trade-off between the expected amount of drift in the system and the acceptable level of uncertainty added to the updated prior. Note that it may be necessary to first train an intermediate flow in order to generate more samples before perturbation, depending on the number of samples originally generated by the Bayesian sampler.

\subsubsection{Inference}
Once a fit has completed, prior or posterior samples can be analyzed, allowing understanding of correlations, insensitivities, and multi-modal behavior. Results can be used for model evaluation, including to compare model physicality; determine parameters that are biased or can be fixed; and understand the influence of noise and error. Inference can be done over functional posteriors, with uncertainties for metrics such as S-parameters or gains quantified using credible intervals. Parameter sensitivities can also be investigated. A method using PAWN indices \cite{pianosi2018distribution} has proven useful in this regard. PAWN allows for a global sensitivity analysis from generic input-output samples. To use the algorithm, weighted samples should first be re-sampled to obtain equally-weighted points. These points are then passed, alongside the function of interest (\textit{e.g.} a frequency-dependent function, perhaps reduced to a lower dimension), to the algorithm. If prior points are used, the output indices are interpreted as representing the function's initial sensitivity to each parameter. However, if posterior points are used, the analysis has been conditioned on the data, and the indices therefore represent ``updated'' sensitivities. Note that, by default, these indices are ``distribution-based'', and do not take into account absolute output variation. To compare sensitivities across datasets, indices can be scaled by the function's variance, allowing for a relative comparison.

\subsubsection{Implementation}
Since circuit models are often highly degenerate in their parameters, care should be taken when selecting a Bayesian sampling algorithm. For example, Markov chain Monte Carlo (MCMC) methods do not deal well with highly correlated or multi-modal posteriors, and also suffer from the curse of dimensionality. For the high-dimensional circuit models in this work, modern algorithms based on nested sampling have been found to work well, specifically PolyChord  \cite{handley2015polychordcosm, handley2015polychord} and BlackJAX \cite{blackjax2020github} for GPU-accelerated model fitting. These samplers are both supported in ParamRF. For PolyChord, the main setting is the number of ``live points'', $n_\textnormal{live}$, usually set to some multiple, $\kappa_\textnormal{live}$, of the parameter count.

\subsection{Modeling and Fitting Pipeline}\label{sec:pipeline}
\begin{figure}[!htb]
    \centering
    \includegraphics[width=0.78\linewidth]{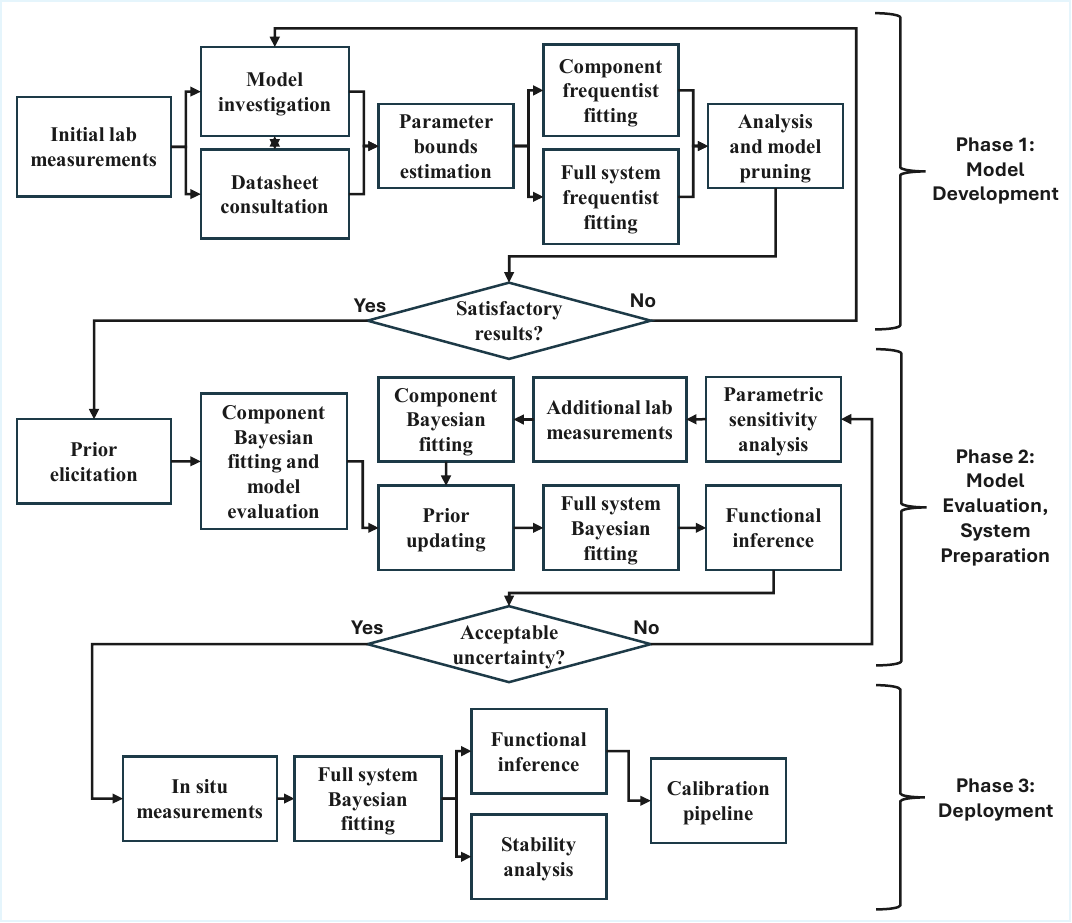}
    \caption{The suggested modeling and fitting pipeline, split into model development (phase 1), model evaluation and system preparation (phase 2), and deployment (phase 3).}
    \label{fig:pipeline}
\end{figure}

Although a large degree of parallelization is possible, full Bayesian fits of such high-dimensional models can still take between hours and days to complete, mainly dependent on the value of $n_\textnormal{live}$. A more practical workflow combining frequentist and Bayesian techniques is therefore presented in Fig.~\ref{fig:pipeline}. Phase 1 involves traditional model development using datasheets and frequentist optimization. Phase 2 then runs Bayesian fitting on the individual components to perform model evaluation, prior updating, and initial inference. Here, it is useful to use a lower $\kappa_\textnormal{live}$ than intended for the final fits, depending on the required level of accuracy of the posterior. For example, experimental fits may use $\kappa_\textnormal{live} = 
1$, whereas final fits should use at least $\kappa_\textnormal{live} = 25$. Finally, phase 3 re-runs the Bayesian fit on the full model, which provides insight into the system and its stability, and would serve as input to the calibration pipeline. In this work, phases 1 and 2 of the pipeline are demonstrated.

\section{Modeling}\label{sec:source_modeling}
This section presents the specific circuit models developed for the case of the REACH calibration sources (phase 1 of the pipeline). In this work, the models for each RF component are physically motivated and grounded in rigorous microwave circuit theory derived from first principles. Compared to less physical models, this approach allows any simplifying assumptions to be identified and investigated if necessary.

\begin{figure}[!htb]
    \centering
    \includegraphics[width=0.75\linewidth]{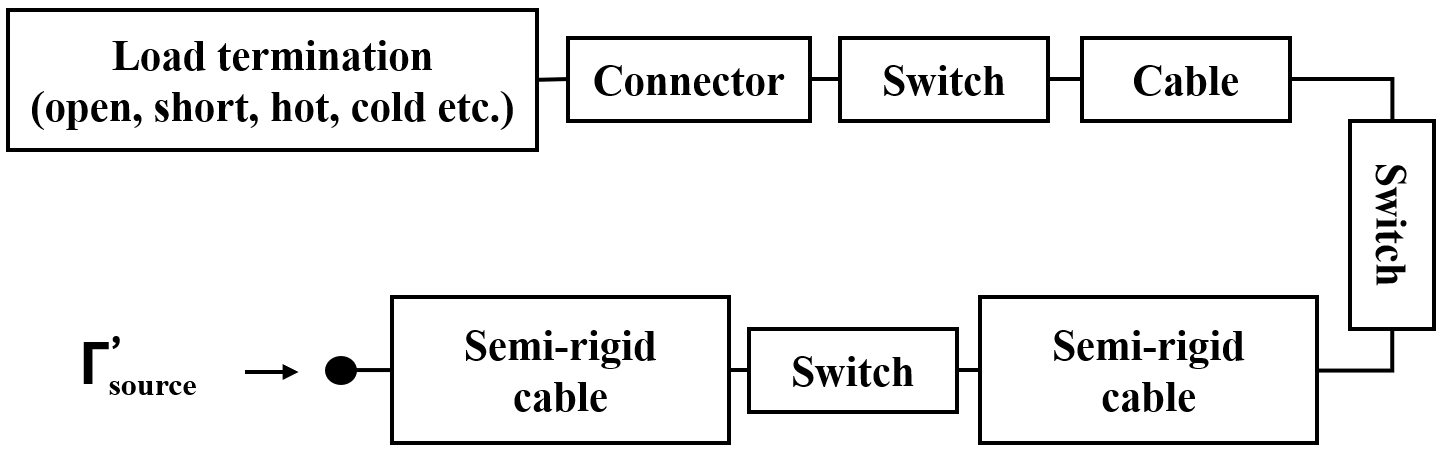}
    \caption{A block diagram for one of the REACH calibrator source chains. $\Gamma'_\mathrm{source}$ is the reflection coefficient being modeled (up until the VNA reference plane in Fig.~\ref{fig:reach_frontend}).}
    \label{fig:source_chain}
\end{figure}

Each source chain can be represented as a cascade of multiple RF components, as illustrated in Fig.~\ref{fig:source_chain}. The flexible cables, semi-rigid cables, switches, and adapters in the chain are all represented using the foundational RLGC transmission line model, whereas the load is modeled using a combination of lumped and transmission line elements. After termination in a load, the resulting source reflection coefficient, $\Gamma_\mathrm{source}'$, is obtained. Here, the prime notation is used to indicate the modeled source reflection coefficient up until the VNA reference plane in Fig.~\ref{fig:reach_frontend} (as opposed to until the main reference plane).

Conventional circuit modeling approaches often apply simplifications in these chains, such as grouping adjacent components into a single equivalent block. In this work, each component is modeled separately for two main reasons. Firstly, since each component can be fit directly to their own laboratory measurements, their resultant parameter posteriors can be used as priors for future fits on the full chain. This can help break degeneracies between components. Secondly, because the exact effect of the combined impedance mismatches and insertion losses is unclear, it is not obvious whether combining adjacent components is appropriate for the level of precision required. Further analysis should therefore be done before making such simplifications. It should be noted that the connectors in the signal chain are indeed grouped with their adjacent components. This simplification is typically appropriate and necessary, since each connector represents only a small length of extra transmission line which also cannot be measured separately. This is not done, however, for the load termination and connector, due to their distinct physical configuration.

\subsection{Coaxial Cables}\label{sec:long_cables}
All cables used in the REACH receiver are coaxial, consisting of a dielectric layer sandwiched between inner and outer conductors. To conform to the uniform RLGC representation, the following assumptions are made:
\begin{arabiclist}
    \item The cables consist of inner and outer conductors with constant radii $a$ and $b$ along their length.
    \item The conductors are effectively solids, with electrical conductivities $\sigma_a$ and $\sigma_b$ for the inner and outer conductors respectively.
    \item The dielectric between these conductors is homogeneous and characterized by a single, frequency-dependent magnetic permeability, $\mu(\omega)$, and electric permittivity, $\epsilon (\omega)$.
\end{arabiclist}
Usually, the permeability is taken to be that of free-space, $\mu(\omega) = \mu_0$, and the permittivity is represented as complex via a relative permeability and loss tangent, $\epsilon(\omega) = \epsilon_0\epsilon_r (\omega)(1 - j\tan\delta(\omega))$. Given these assumptions, and assuming that the ``skin effect'' is fully present, a field analysis can be conducted, as summarized in \citet{tesche2007simple}, resulting in a ``physical'' coaxial model,
\begin{align}
    R'(\omega) &= \frac{\sqrt{\mu_0}}{2 \pi} \left(\frac{1} {a\sqrt{2\sigma_a}} + \frac{1}{b\sqrt{2\sigma_b}}\right) \cdot \sqrt{\omega},    
    \label{eqn:RLGC_R} \\
    L'(\omega) &= \frac{\mu_0}{2 \pi} \ln{\frac{b}{a}} + \frac{\sqrt{\mu_0}}{2 \pi} \left(\frac{1}{a\sqrt{2\sigma_a}} + \frac{1}{b\sqrt{2\sigma_b}}\right) \cdot \frac{1}{\sqrt{\omega}},    
    \label{eqn:RLGC_L} \\
    G'(\omega) &= \frac{2 \pi \epsilon_0}{\ln b/a}\cdot \epsilon_r(\omega)\tan\delta(\omega) \cdot\omega,
    \label{eqn:RLGC_G} \\
    C'(\omega) &= \frac{2 \pi \epsilon_0}{\ln{b/a}} \cdot \epsilon_r(\omega).
    \label{eqn:RLGC_C}
\end{align}

\subsection{Flexible Cables}
A number of assumptions for (\ref{eqn:RLGC_R})-(\ref{eqn:RLGC_C}) have proven appropriate for the flexible $\SI{2}{m}$ and $\SI{10}{m}$ cables in REACH \cite{TimesMicrowave_cable}. Firstly, since both conductors are made of copper, $\sigma_a$ and $\sigma_b$ are both set equal to a new conductivity, $\sigma_c$. Secondly, it is noted that the forms of the dielectric constant and loss tangent, $\epsilon_r(\omega)$ and $\tan\delta(\omega)$, are still general. Several wideband dispersion models exist, such as the Debye model \cite{debye1913theorie}, which model the material parameters by directly describing the particle relaxation effects. However, these models are often tailored towards wideband, higher-frequency scenarios, and consequently have a larger number of parameters and complexity. Further, the dispersive effect of the materials of interest within the REACH frequency band, specifically polyethylene, are generally slowly varying with frequency. Practically, a constant value across frequency has proven to work well. However, to allow more detailed model comparison, and cater for future, more dispersive materials, a general $n$th order polynomial model is assumed. In this case, a good approach is to use the Bernstein basis, since this allows the bounds or priors for each coefficient to be set to the same expected value. In the linear case ($n=1$), and written in terms of normalized frequency, $\omega_n$,
\begin{align}\label{eqn:dielectric_parameters}
    \epsilon_r(\omega_n) &\approx \epsilon_{r_0}(1-\omega_n)  + \epsilon_{r_1}\omega_n, \\
    \tan\delta(\omega_n) &\approx \tan\delta_0(1-\omega_n) + \tan\delta_1 \omega_n.
\end{align}

\subsection{Semi-ridged Cables and Switches}
The semi-rigid cables and switches in REACH are much shorter than the main long cables (\textit{e.g.} $\SI{10}{cm}$ for the hot load cable), and are also all electrically short. For such lengths, the effect of dispersion becomes negligible, and it is therefore assumed that the material properties can be represented as constant across frequency. Also, the datasheets for such cables typically quote a velocity factor, $v_f = \frac{1}{\sqrt{\epsilon_r}}$, and attenuation factors, $k_1$ and $k_2$, instead of material properties. Therefore, the RLGC parameters are re-written in terms of these factors, as well as a nominal characteristic impedance, $Z_n$, resulting in a ``datasheet'' model,
\begin{align}
    R'(\omega) &= 2 Z_n \frac{\ln10}{20} k_1 \sqrt{\omega}, \nonumber &
    L'(\omega) &= \frac{Z_n}{v_f c_0} + \frac{R'(\omega)}\omega,  \nonumber \\
    G'(\omega) &= \frac{2}{Z_n} \frac{\ln{10}}{20} k_2 \omega, &
    C'(\omega) &= \frac{1}{Z_n v_f c_0}, \label{eqn:RLGC_datasheet}
\end{align}
where $c_0$ is the speed of light in a vacuum. Note that $k_1$ represents skin effect/conductive losses, and $k_2$ dielectric losses, with the total cable attenuation being proportional to $k_1 \sqrt{\omega} + k_2 \omega$.

There is a high level of uncertainty regarding the internals of the switches. Isolated two-port measurements have revealed several unique characteristics and complexities that require further investigation if to be treated fully. However, extensive testing has ultimately shown that they can be approximated as short, well-matched transmission lines in the context of the full signal chain. In this work, they will therefore assume the datasheet model in (\ref{eqn:RLGC_datasheet}). Since their internal dielectric constant and length are unknown, it will be assumed that $v_f=1$, meaning that their length in the model is an ``effective'' one. Further, it will be assumed that their dielectric losses are zero ($k_2=0$), which is typically valid at these frequencies for high-quality switches and connectors, but may be accounted for if necessary.

\subsection{Loads}\label{sec:source_modeling_loads}
\begin{figure}[!htb]
    \centering
    \includegraphics[width=0.9\linewidth]{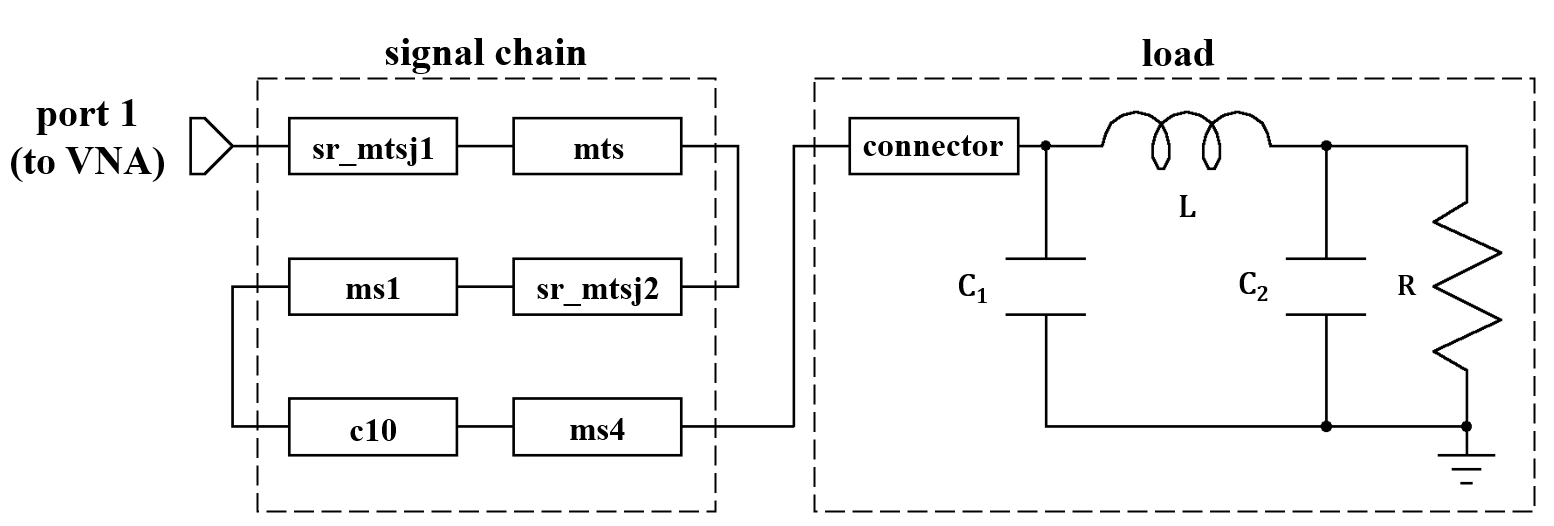}
    \caption{A circuit model for the c10r10 source at the VNA reference plane using the single line load model. The load connector, switches (mts, ms1, ms4) and semi-rigid coaxials (sr\_mtsj1, sr\_mtsj2) use the simplified datasheet model, with $k_2=0$ for connectors and switches. The flexible $\SI{10}{m}$ cable (c10) uses the full physical coaxial model. Sources are named according to their cable and resistor, and semi-rigids according to their connecting switch \textit{e.g.} sr\_mtsj1 which connects to J1 of MTS (see Fig.~\ref{fig:reach_frontend}).}
    \label{fig:source_model}
\end{figure}

The matched loads in REACH (\textit{i.e.} hot and cold) consist purely of off-the-shelf-components, whereas the non-matched loads (\textit{e.g.} $\SI{10}{\ohm}$ or $\SI{91}{\ohm}$) are custom manufactured. For these custom loads, each consists of a SubMiniature version A (SMA) and male-to-male (MTM) connector, with one or more resistors soldered across the SMA's end. Since the resistor is in an open-ended configuration, there will be some parasitic inductance in its legs, and parasitic capacitance with its connector. This is usually modeled using two capacitors, $C_1$ and $C_2$, and one inductor, $L$, in a $\pi$-CLC topology, with the ABCD matrix
\begin{equation}\label{eqn:abcd_clc}
    \mathbf{A}_{\textnormal{CLC}} =
    \begin{bmatrix}
        1 + \frac{Y_2}{Y_3}                         & \frac{1}{Y_3} \\
        Y_1 + Y_2 + \frac{Y_1 Y_2}{Y_3}             & 1 + \frac{Y_1}{Y_3} \\
    \end{bmatrix},
\end{equation}
where $Y_1 = j\omega C_1$, $Y_2 = j\omega C_2$, and $Y_3 = \frac{1}{j\omega L}$. This $\pi$-CLC topology is shown in Fig.~\ref{fig:source_model}.

The load connectors will also exhibit some conductive losses and phase offset, however the relevance of modeling these explicitly depends on the reference plane taken for the gain in the temperature correction. To enable an in-depth model comparison, three models are considered, namely 1) a ``$\pi$-CLC'' model, catering for the parasitic edge effects with a $\pi$-CLC network but not modeling the connector effects, 2) a ``single line'' model, with one short transmission line appended to the $\pi$-CLC model representing both connectors as one, and 3) a ``double line'' model, similar to the single line model but with two transmission lines catering for each connector separately. An example of a full model for the c10r10 source using the single line model and with all components named is shown in Fig.~\ref{fig:source_model}.

\subsection{Full Chain}

\begin{table}
    \centering
    \caption{Model Parameters Before Simplification}
    \begin{tabular}{lrlr}
    \hline
        \textbf{Model} & \textbf{No.}  & \textbf{Parameters}  & \textbf{Total} \\ \hline
        Loads & 12 & $R$, $C_1$, $L$, $C_2$, $\ell$, $Z_n$, $k_1$ & 84  \\ 
        Flexible Cables & 2 & $\ell$, $a$, $b$, $\sigma_c$, $\epsilon_{r_0}$, $\tan\delta_0$ & 12  \\ 
        Semi-ridged Cables & 3 & $\ell$, $Z_n$, $v_f$, $k_1$, $k_2$ & 15  \\ 
        Switches & 4 & $\ell$, $Z_n$, $k_1$ & 12  \\ \hline
        \textbf{Total} & & & \textbf{123}  \\ \hline
    \end{tabular}
    \vspace{0.2cm}
    \caption*{The distribution of the model parameter space before further simplification. A constant model is assumed for the flexible cable material parameters, and the single line model is used for the loads. $v_f$ is fixed to one for the switches and $\frac{1}{\sqrt{2.1}}$ for the load connectors (which are typically polytetrafluoroethylene (PTFE)). The final model reflection coefficient, $\Gamma'_\mathrm{source}$, is found to be most sensitive to the load resistance $R$ and flexible cable parameters such as $\ell$, $\epsilon_{r_0}$, $Z_n$, $a$ and $b$. It is found to be least sensitive to loss parameters, such as $k_1$, $k_2$, $\sigma_c$ and $\tan\delta_0$.}
    \label{tab:parameters}
\end{table}

By multiplying the ABCD matrices of the individual components in the source chain up until the resistor, a matrix representing the full cascaded signal chain is obtained. To obtain the chain's S-parameters, $S_{11}$ and $S_{21}$, the cascaded matrix can be converted to S-parameters using common conversion formulae. To obtain the combined reflection coefficient for fitting, $\Gamma'_\mathrm{source}$, the matrix must be terminated in the resistive load. For a load resistance $R$, cascaded chain parameters $A$, $B$, $C$ and $D$, and reference impedance $Z_0$,
\begin{equation}\label{eqn:chain_s11}
    \Gamma'_{\textnormal{source}} =
    \frac{Z_0 (1 + \Gamma_{R})(A-Z_0 C) + (B - D Z_0) (1 - \Gamma_{R})}{Z_0 (1 + \Gamma_{R})(A+Z_0 C) + (B + D Z_0) (1 - \Gamma_{R})}
\end{equation}
where $\Gamma_R = \frac{R-Z_0}{R+Z_0}$. Across all source chains and components, this results in a total of 123 parameters before further model simplifications. A summary of the final parameter space is shown in Table \ref{tab:parameters}.

\begin{figure}[!htb]
    \centering
    \includegraphics[width=0.85\linewidth]{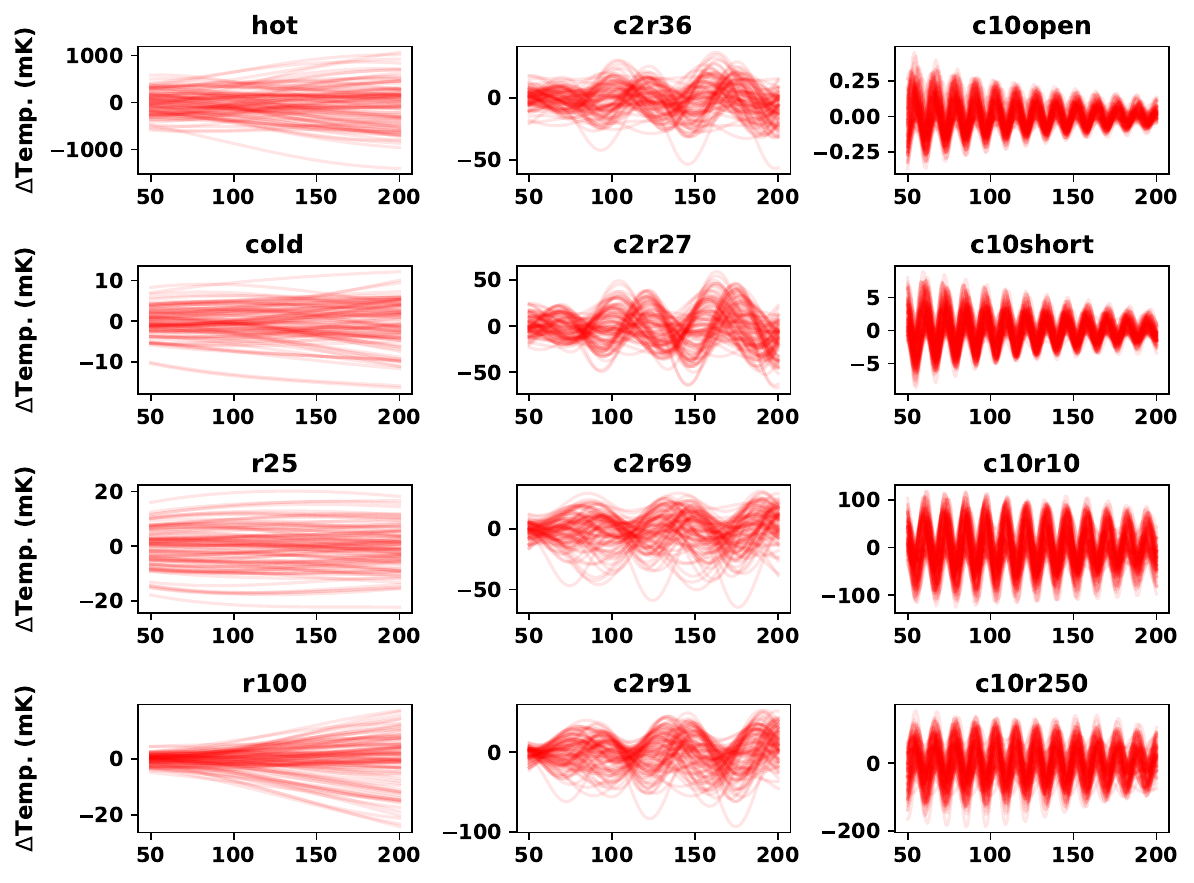}
    \caption{Effective source temperature differences from nominal values, calculated for all 12 independent calibrator chains using (\ref{eqn:temperature_corrections}). Parameter samples are drawn from a 10\% uniform distribution of all parameters in Table \ref{tab:parameters}. Physical temperatures used are $T_\mathrm{cab} = \SI{298}{K}$, $T_{R_\mathrm{hot}} = \SI{370}{K}$ and $T_{R_\mathrm{other}} = \SI{297}{K}$. The $x$ axis represents frequency in MHz. The graph highlights the sensitivity of the effective temperatures to changes in model parameters, which can vary up to 1 K depending on the specific source and amount of deviation.}
    \label{fig:temperature_sensitivities}
\end{figure}

As an initial investigation, the model's parameters were varied about nominal datasheet values and the change in the resulting effective source temperatures plotted using (\ref{eqn:temperature_corrections}). The results in Fig.~\ref{fig:temperature_sensitivities} show temperature differences on the order of \SI{100}{mK} to \SI{1000}{mK} (depending on the source) for a 10\% variation in the system parameters. This highlights the system's sensitivity to relatively small perturbations.

\graphicspath{{figures/}}

\begin{figure}[!htb]
    \centering
    \includegraphics[width=0.95\columnwidth]{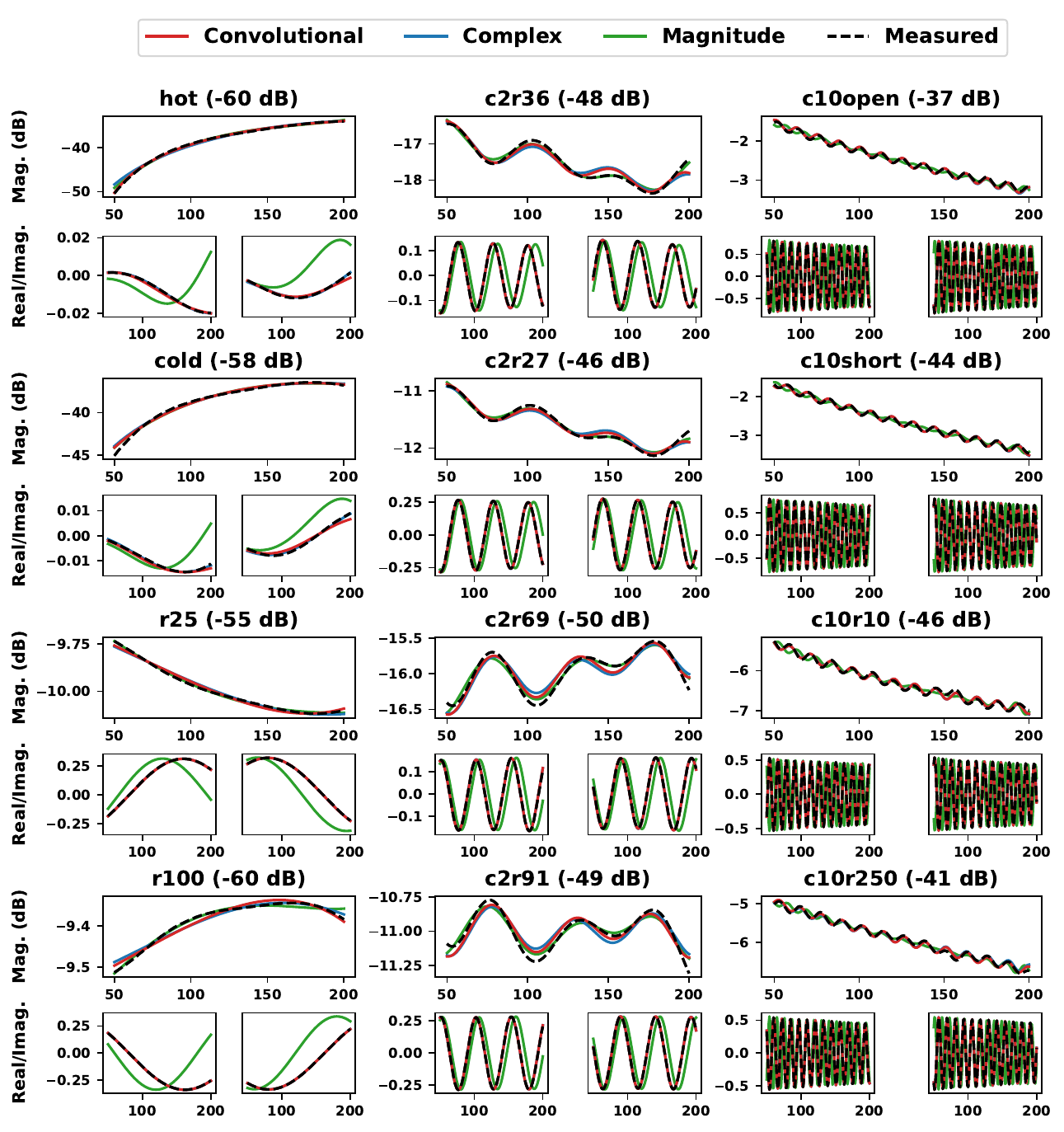}
    \caption{Comparison of the model vs. measured $\Gamma'_\mathrm{source}$ for all 12 independent calibrators optimized simultaneously using (\ref{eqn:conv_cost}) for $\mathcal{D}_\mathrm{asm}$. $x$ axis values are in MHz. Root-mean-square residuals for the convolutional cost are shown in brackets. The results obtained using the convolutional cost emphasize the accuracy of the model, exhibiting a worst-case RMS error of -37 dB.}
    \label{fig:frequentist_fit}
\end{figure}
\section{Results}\label{sec:results}
This section presents the results for the REACH sources, which includes the best fits of the models, parameters estimated using Bayesian inference, model comparisons and evaluations, and detailed analysis for the motivating case of the temperature corrections (phases 1 and 2 of the pipeline).

\clearpage
In order to demonstrate all the proposed techniques effectively, three datasets are used:
\begin{itemlist}
    \item $\mathcal{D}_\mathrm{lab}$: Individual, two-port laboratory measurements of the RF components in the source signal chains, as well as one-port measurements of the loads with their connectors.
    \item $\mathcal{D}_\mathrm{asm}$: Full, in situ one-port measurements of the assembled system's source reflection coefficients.
    \item $\mathcal{D}_\mathrm{sim}$: A circuit simulation of a ``reduced'' source model, consisting of a signal chain modeled using a datasheet coaxial cable as in (\ref{eqn:RLGC_datasheet}) and a $\pi$-CLC load. Only a representative, non-ideal hot source with parasitics is simulated. Both a ``nominal'' and a ``perturbed'' simulation are used. Nominal parameters are $R = \SI{45}{\ohm}$, $\ell = \SI{500}{mm}$, $Z_n = \SI{55}{\ohm}$, $v_f = 1$, $k_1 = \SI{1}{dB}/(\SI{100}{m}\,\cdot\sqrt{\text{MHz}})$, $k_2 = \SI{0.001}{dB}/(\SI{100}{m}\,\cdot\text{MHz})$, $C_1 = \SI{1}{pF}$, $L = \SI{1}{nH}$ and $C_2 = \SI{1}{pF}$. Perturbed parameters have the signal chain's $k_1$ loss increased by 5\%. Gaussian measurement noise is added, with $\sigma = 10^{-4}$.
\end{itemlist}

All results are presented in the order of the pipeline in Fig.~\ref{fig:pipeline}. Frequentist fits were run using a combination of SLSQP and Nelder-Mead based optimization. Bayesian fits were run using PolyChord with $\kappa_\textnormal{live} = 25$ unless otherwise stated. Either informative, uninformative, or MAF-based, updated priors were used. Informative and uninformative priors are listed in Tables \ref{tab:load_priors} to \ref{tab:switch_priors}. In all cases, the data was interpolated onto 151 frequency points between $50$ and $\SI{200}{MHz}$.

\subsection{Joint Fit}
Results for the joint fit of all 12 sources using $\mathcal{D}_\mathrm{asm}$ and frequentist optimization are presented in Fig.~\ref{fig:frequentist_fit}. Fits are done using convolutional, complex, and magnitude cost functions before any model simplification (\textit{i.e.} using the parameters as in Table \ref{tab:parameters}). For the convolutional cost, an RMS error of no worse than $\SI{-37}{dB}$ is obtained for the complex residuals across all sources, which is considered a strong agreement in the context of circuit model fitting. Magnitudes plotted on a log scale also match well visually, exhibiting only slight deviations near the ends of the frequency band. It is noted that a systematic residual is found in the $\SI{2}{m}$ and $\SI{10}{m}$ sources, potentially caused by unmodeled cable deformations. Overall, however, the results demonstrate how the complex and magnitude error functions can provide unsatisfactory results for several sources, while the convolutional cost provides a compromise between both metrics, resulting in a good fit for all magnitude and real/imaginary values.

\subsection{Model Evaluation}
To understand the degeneracies and biases in the models, posteriors of individual components in $\mathcal{D}_\mathrm{lab}$ were estimated using Bayesian inference and uninformative priors. Since the goal in REACH is to utilize one-port measurements to infer general two-port properties, it is useful to conduct some two-port fits using reflection and transmission coefficient data separately, in order to fully understand how each of these measurements informs the parameters. Given that the hot source has the greatest temperature uncertainty, its 10 cm semi-rigid cable is used as a case study to evaluate the cable models, while the $\SI{91}{\ohm}$ custom load is used to evaluate the load models.

\begin{figure}[!htb]
    \centering
    \includegraphics[width=0.52\columnwidth]{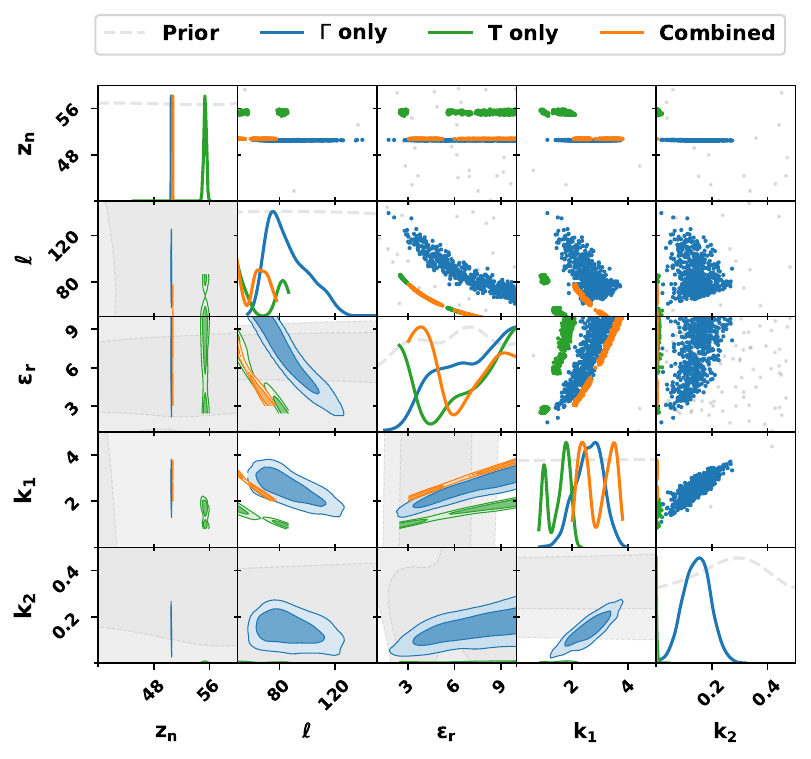}
    \caption{Parameter posteriors of the $\SI{10}{cm}$ semi-rigid cable in the hot source, fit using uninformative priors and two-port data. 1D and 2D marginal posteriors are shown, estimated using Gaussian kernel density estimation (KDE). The results highlight how reflection-only data can bias estimates for parameter combinations such as $\epsilon_r$ - $\ell$, or for loss parameters such as $k_1$ and $k_2$. Further, the results imply that good priors on $\epsilon_r$ and $\ell$ are needed, due to their correlations with other parameters.}
    \label{fig:sr_ms1j2_parameter_posteriors}
\end{figure}
\begin{figure}[!htb]
    \centering
    \includegraphics[width=0.58\columnwidth]{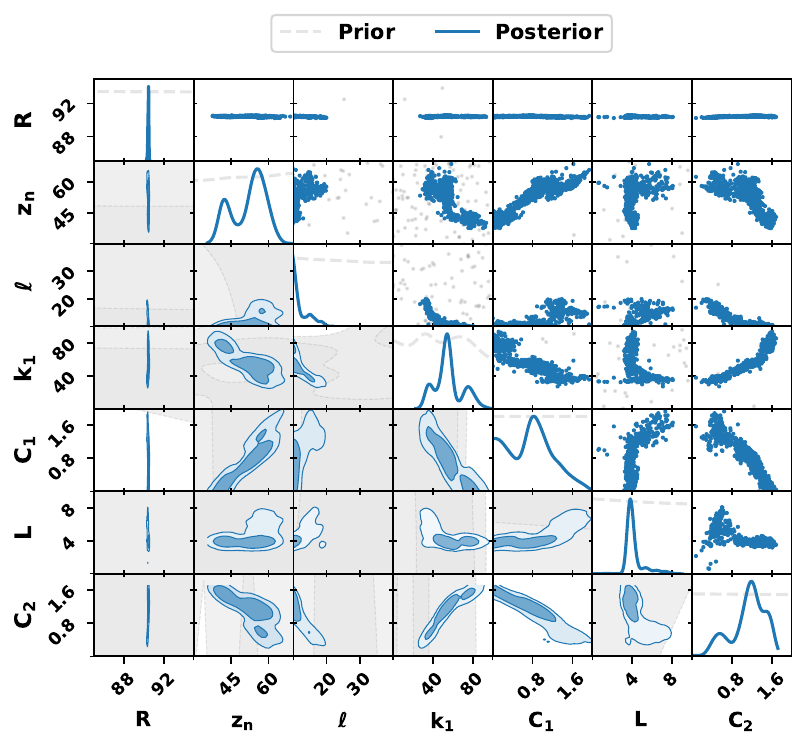}
    \caption{Parameter posteriors of the $\SI{91}{\ohm}$ load, fit using uninformative priors, the single line model, and one-port data. Marginal posteriors are plotted as in Fig.~\ref{fig:sr_ms1j2_parameter_posteriors}. The results demonstrate the complex relationships between parameters, as well as the need for informative priors if the load is to be disentangled.}
    \label{fig:r91_load_parameter_posteriors}
\end{figure}

The semi-rigid cable posteriors are shown in Fig.~\ref{fig:sr_ms1j2_parameter_posteriors}. It is clear that there is a large curving degeneracy between the cable length, $\ell$, and the dielectric constant, $\epsilon_r$, with different curves representing different electrical lengths. Further, the $\epsilon_r$ - $\ell$ curves for the reflection and transmission coefficients datasets do not align, however the combined fit almost entirely favors the transmission coefficient data. This shows how the transmission coefficient, $T$, provides an informative estimate of the electrical length, however that the estimate provided by the reflection coefficient, $\Gamma$, is biased, likely due to a combination of model error and measurement noise. Similarly, the losses, $k_1$ and $k_2$, are seen to be highly informed by $T$ and only slightly informed by $\Gamma$, whereas $Z_n$ is highly informed by $\Gamma$ and slightly misinformed by $T$.

For the $\SI{91}{\ohm}$ load, it is found that the parasitic capacitances, $C_1$, and $C_2$, are strongly correlated with each other, but not with $L$. $R$ appears to be relatively well-informed by the data, however the line parameters, $z_n$ and $\ell$, have noticeable uncertainties. The connector loss, $k_1$, appears to be almost entirely uninformed. This follows intuitively from the previous analysis. Therefore, if the load model is to be disentangled (\textit{e.g.} for a connector temperature correction), then highly informative priors will be needed.

\subsection{Model Comparison}
In order to compare and simplify the models where possible, Bayesian evidences were calculated of various component-level fits on $\mathcal{D}_\mathrm{lab}$ with different parameters fixed. Unless otherwise stated, informative priors were used.

For the semi-rigid cables, it is found that the velocity factors (or dielectric constants) may be fixed, but that no further simplifications can be made. For the longer, flexible cables, it is found that both the dielectric constant and either of the shield diameters may be fixed, but only if uninformative priors are used. However, if informative priors are used (even with bounds that are much looser than the expected physical tolerances) the evidence no longer permits any parameters to be fixed. In other words, simplifications that seem acceptable under uninformative priors do not hold once the parameter space is constrained to realistic values. This suggests that the model is relying on the full flexibility of its parameter space in a way that is not fully physical, which implies that the physicality of the model itself could be improved. To this end, no simplifications will be made for these cables.

For the custom, resistor-based loads, since they contribute the largest number of parameters, a more in-depth comparison was done. The resultant evidence table is shown in Table \ref{tab:load_evidences}, from which a number of conclusions can be drawn:
\begin{arabiclist}
    \item Although the $\pi$-CLC model performs better than some simpler line models, it alone is not sufficient \textit{i.e.} a lossy, mismatched line is indeed necessary.
    \item Parasitics capacitances and inductances should be included, however a single parasitic by itself does not have a large impact ($D_3$, $D_4$, $D_5$ and $D_6$). The best parasitic combination consists of $C_1$ and $L$ \textit{i.e.} $C_2$ may be fixed to zero ($D_{7}$ vs. $D_{9}$ and $S_{3}$ vs. $S_{4}$).
    \item In general, it is not necessary to model the SMA and MTM adapters as two separate transmission lines ($S_4$ vs. $D_{9}$ and $S_{3}$ vs. $D_{7}$).
\end{arabiclist}

\begin{table}[!htb]
    \centering
    \caption{Load Model Evidences}
    \begin{tabular}{c|c|c|c|c|c|c|c}
    \hline
    \textbf{Model} & $\ell$ & $Z_n$ & $k_1$ & $C_1$ & $L$ & $C_2$ & \textbf{Log. Evidence} \\ \hline
    $D_{1}$ &       & $\checkmark$ &       &       &       &       & 13468 \\ \hline
    $D_{0}$ & $\checkmark$ &       &       &       &       &       & 14184 \\ \hline
    $D_{2}$ & $\checkmark$ & $\checkmark$ &       &       &       &       & 15127 \\ \hline
    $P_{0}$ &       &       &       & $\checkmark$ & $\checkmark$ & $\checkmark$ & 15140 \\ \hline
    $S_{1}$ & $\checkmark$ & $\checkmark$ & $\checkmark$ & $\checkmark$ &       &       & 15825 \\ \hline
    $S_{2}$ & $\checkmark$ & $\checkmark$ & $\checkmark$ &       & $\checkmark$ &       & 15828 \\ \hline
    $S_{0}$ & $\checkmark$ & $\checkmark$ & $\checkmark$ &       &       &       & 15847 \\ \hline
    $D_{3}$ & $\checkmark$ & $\checkmark$ & $\checkmark$ &       &       &       & 15857 \\ \hline
    $D_{4}$ & $\checkmark$ & $\checkmark$ & $\checkmark$ & $\checkmark$ &       &       & 15862 \\ \hline
    $D_{5}$ & $\checkmark$ & $\checkmark$ & $\checkmark$ &       & $\checkmark$ &       & 15877 \\ \hline
    $D_{6}$ & $\checkmark$ & $\checkmark$ & $\checkmark$ &       &       & $\checkmark$ & 15880 \\ \hline
    $D_{8}$ & $\checkmark$ & $\checkmark$ & $\checkmark$ &       & $\checkmark$ & $\checkmark$ & 15893 \\ \hline
    $D_{9}$ & $\checkmark$ & $\checkmark$ & $\checkmark$ & $\checkmark$ & $\checkmark$ & $\checkmark$ & 16311 \\ \hline
    $D_{7}$ & $\checkmark$ & $\checkmark$ & $\checkmark$ & $\checkmark$ & $\checkmark$ &       & 16336 \\ \hline
    $S_{4}$ & $\checkmark$ & $\checkmark$ & $\checkmark$ & $\checkmark$ & $\checkmark$ & $\checkmark$ & 16348 \\ \hline
    $S_{3}$ & $\checkmark$ & $\checkmark$ & $\checkmark$ & $\checkmark$ & $\checkmark$ &       & 16354 \\ \hline
    \end{tabular}
    \vspace{0.2cm}
    \caption*{An evidence table comparing the various load models. The evidences are summed for all the custom, resistor-based loads which share the same physical configuration  (\textit{i.e.} all loads except hot, cold, open and short). $P_i$, $S_i$ and $D_i$ refer to the $\pi$, single, and double line models respectively. Free parameters are indicated with ticks. $k_2$ is fixed to zero in all cases. From the table, it can be deduced that $C_2$ should be fixed to zero, and that only a single transmission line is needed.}
    \label{tab:load_evidences}
\end{table}

The final model for the custom loads is therefore chosen as a single line with a CL network ($S_3$). For the hot and cold loads, however, it is unclear whether such parasitics are necessary, since typical off-the-shelf matched loads are specifically designed to minimize these effects. A separate comparison was therefore done on these loads using models $S_0$ to $S_3$. From these results, it is concluded that $S_0$ is sufficient for the matched loads \textit{i.e.} no parasitics are required.

Overall, the above simplifications reduce the parameter space in Table \ref{tab:parameters} from 123 to 104 parameters. This will reduce future computation time, help prevent over-fitting, avoid unnecessary degeneracies in the parameter space, and simplify any further analysis.

\subsection{Temperature Uncertainties}
Although the previous sections provide insight into parameter uncertainties and model biases, it is initially unclear how these uncertainties translate to the final predicted temperatures. Given that the global 21 cm signal has an estimated depth on the order of a few hundred mK, such uncertainties must be lowered to a similar order of magnitude. This section therefore investigates the source temperatures directly, as calculated using (\ref{eqn:temperature_corrections}), with all parameters predicted by reflection coefficient fits. Frequentist fits are run with either the complex or convolutional cost functions, and Bayesian fits are run with either one-dimensional, ``datasheet'' priors, or MAF-trained, multi-dimensional, ``updated'' priors. The hot source is used as a case study, and the results obtained using both simulated ($D_\mathrm{sim}$) and measured ($D_\mathrm{asm}$) data is analyzed. 

\subsubsection{Simulated}
For the simulated case, fits are run using both frequentist and Bayesian approaches. First, component-level fits are run on the nominal cable and load with relatively wide datasheet (uniform) priors to simulate the laboratory measurement process. For frequentist fits, the resultant parameters are used as the average for updated optimizer bounds. For Bayesian fits, the resultant posteriors are used to train MAF networks (with broadening) to be used as updated priors. Using these updated bounds/priors, full source fits are then run on both the nominal and perturbed sources, which are then used to predict the temperature corrections.

\begin{figure}[!htb]
    \centering
    \includegraphics[width=0.7\linewidth]{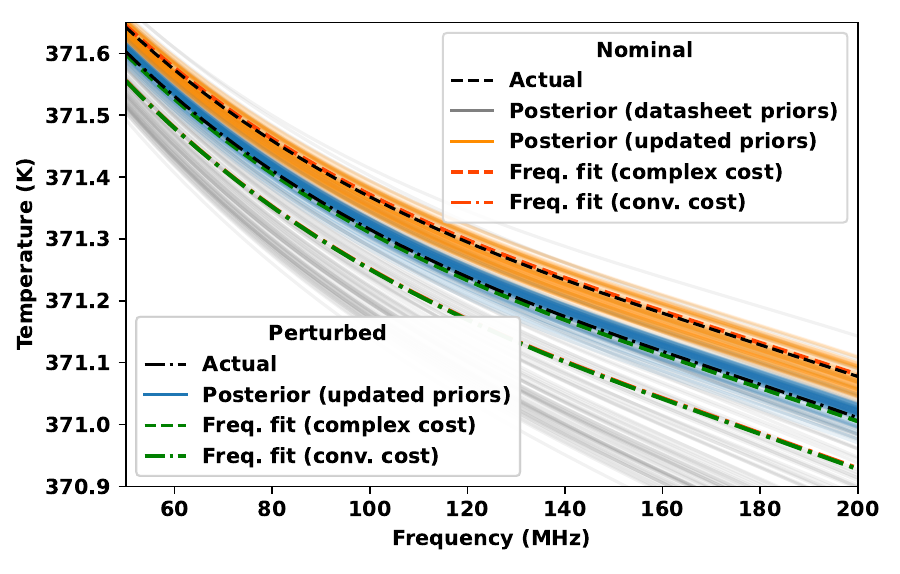}
    \caption{Hot source effective temperature posteriors for the simulated data, $D_\mathrm{sim}$. Temperatures are calculated using (\ref{eqn:temperature_corrections}), with the load (including its connector) as the noise generator, and the end of the cable as the reference plane. The results show how specially-designed cost functions can bias predictions for frequentist fits. Further, it is clear that predictions can have uncertainties on the order of tens to hundreds of mK, depending on the prior and amount of measurement noise.}
    \label{fig:hot_temperatures_simulated}
\end{figure}

The results are shown in Fig.~\ref{fig:hot_temperatures_simulated}. The analysis demonstrates that, even in the case of a perfect model, noise and degeneracies can cause predictions to have temperature errors/uncertainties on the order of hundreds of millikelvin if one-dimensional, datasheet priors are used. This is the case even though the resultant $\Gamma_\mathrm{source}'$ models all fit within $\pm\SI{0.0005}{dB}$ of the true value (not shown). Effectively, the frequentist optimization may converge to any one of the minima illustrated in grey in the figure, depending on the cost function, starting point, and optimization algorithm used. However, such uncertainties can successfully be reduced using updated priors. If samples are broadened to cater for drift, uncertainties on the order of tens of mK can be achieved. Further, if samples are not broadened, the final uncertainty is directly influenced by the amount of measurement noise. For the simulated noise value of $\sigma=10^{-4}$, uncertainties on the order of a few mK are achievable.

\subsubsection{Measured}
For the measured case, only a Bayesian approach is used. Fits are run using the datasheet (informed) priors, as well as priors updated from laboratory measurements of individual components (without broadening). Although the focus of this work is on phases 1 and 2 of the pipeline, a preliminary joint fit is also run using datasheet priors, in order to analyze the effect of including other sources on the hot source temperature predictions.

\begin{figure}[!htb]
    \centering
    \includegraphics[width=0.7\linewidth]{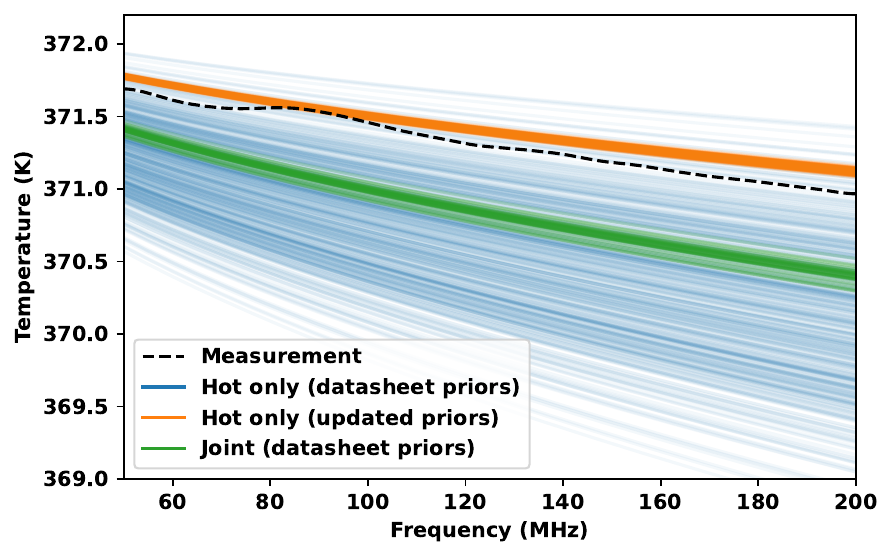}
    \caption{Hot source effective temperature posteriors for the measured data, $D_\mathrm{asm}$. Temperatures are calculated as in the simulated case, but to the main reference plane in Fig.~\ref{fig:reach_frontend}. The ``measurement'' curve shows the conventional correction, which is calculated using a one-port measurement of the hot load, and a two-port measurement of its signal chain. The joint fit was run using $\kappa_\textnormal{live} = 1$. The results shown how datasheet priors can exhibit high uncertainties on the order of $1$ to $\SI{2}{K}$. However, priors updated from laboratory measurements are shown to reduce these uncertainties down to around $\SI{75}{mK}$, and also provide predictions that more closely align with the conventional correction.}
    \label{fig:hot_temperatures_measured}
\end{figure}

The results are shown in Fig.~\ref{fig:hot_temperatures_measured}. The fit using datasheet priors is shown to produce an unacceptable uncertainty on the order of $1$ to $\SI{2}{K}$. However, by using updated priors, this uncertainty is reduced to around $\SI{75}{mK}$. This shows how previous laboratory measurements must be used to combat degeneracies. The prior-updated predictions are also found to align well with the measurement-based correction, with the added advantage of being fit to the in situ measurement of the assembled receiver. Further, since the model is physically-motivated, it has effectively compensated for the VNA systematics.

\clearpage
Compared to the hot only fit with datasheet priors, the joint fit predictions are shown to have a much lower uncertainty of around $\SI{250}{mK}$. This shows how fitting multiple sources simultaneously does reduce some uncertainties in the signal chains, but that this is not enough to pull apart shared intra-chain degeneracies. For example, such a fit will likely not reduce uncertainties in the path between the VNA reference plane and the main reference plane (see Fig.~\ref{fig:reach_frontend}). Further, the joint fit predictions appear to be biased by a similar order of magnitude with respect to the updated and measurement-based temperatures. This shows that biases present in the other source models can substantially affect the hot source predictions. After more detailed analysis, it appears that such biases may be due to the uncaptured non-idealities in the flexible cable models. Therefore, if a joint fit is to be used, further investigation is needed in order to be confident in the resulting temperature predictions.

\subsection{Temperature Sensitivities}
\begin{figure}[!htb]
  \centering
  \includegraphics[width=0.8\columnwidth]{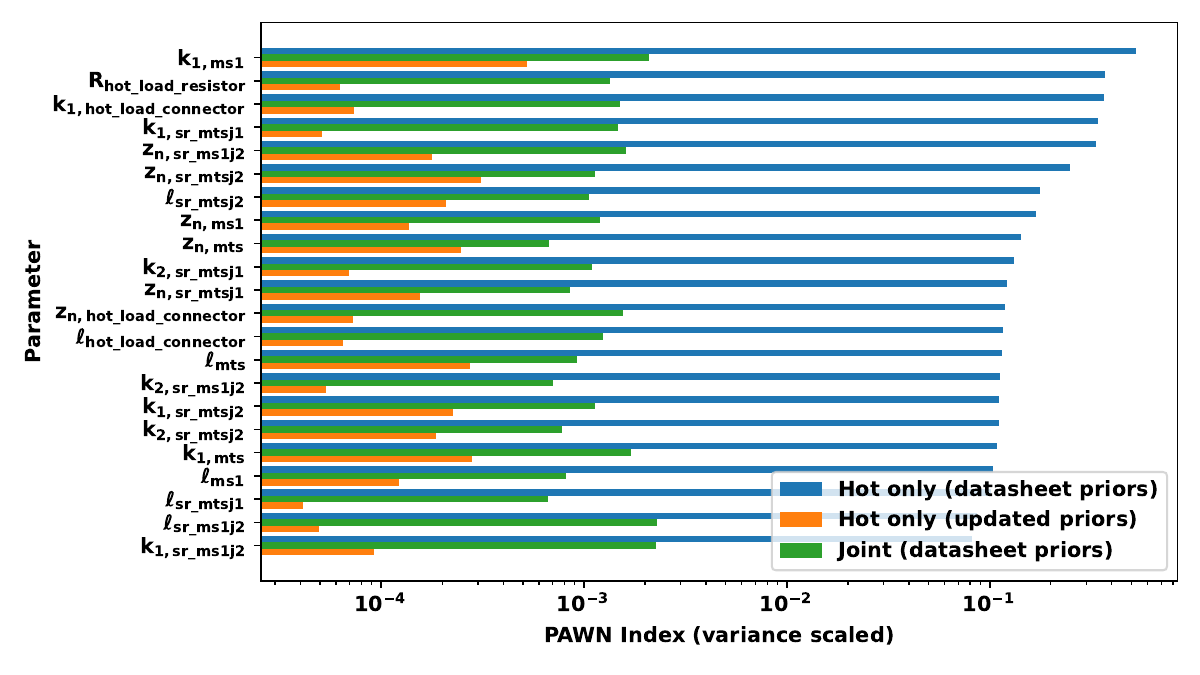}
  \caption{Hot temperature parameter sensitivities computed via PAWN indices for the measured data, $D_\mathrm{asm}$. Indices are sorted by the datasheet prior results, scaled by the output temperature variance, and plotted on a log scale. Semi-rigids are named according to their connecting switch (as described previously). Note that even though mts and sr\_mtsj1 are not used directly in the temperature correction, they still contribute to the overall uncertainty due to parameter correlations. The graph shows how the joint fit reduces sensitivity in shared chain parameters the most, while the updated fit has larger sensitivities in the switches.}
  \label{fig:hot_temperature_indices}
\end{figure}

Parameter sensitivities calculated for the measured hot source using scaled PAWN indices (as described in Section \ref{sec:bayesian_inference}) are shown in Fig.~\ref{fig:hot_temperature_indices}. The indices provide an indication of the relative influence of the model parameters on the hot source temperature predictions, which can be used to prioritize future measurements and modeling. The results show that the joint fit most strongly reduces the temperature sensitivity of parameters that are shared between sources, while the hot source semi-rigid cable exhibits the largest sensitivities. The updated priors, however, show their largest sensitivity in the switch parameters, which may be due to non-idealities that are difficult to capture with the transmission line model.

Note that the magnitude of the datasheet sensitivities are essentially directly controlled by the width of the informative prior. For example, a narrower prior on any skin effect losses, $k_1$, will narrow its posterior and therefore reduce its overall contribution to the temperature uncertainty. However, due to correlations in the parameter space, it may be difficult to narrow such priors around a single value. In this sense, it is safer to place a broader informative prior over a specific parameter, and allowing the prior updating process to reduce the uncertainty (and therefore sensitivity) of a parameter appropriately.

\section{Conclusion}\label{sec:conclusion}
We have introduced a set of modeling, fitting and analysis techniques for in situ monitoring and corrections in global 21 cm experiments, exemplified by temperature corrections for the REACH project. By developing detailed circuit models of the instrument, and applying a combination of frequentist optimization and Bayesian inference, real-time predictions of otherwise immeasurable system parameters can be made. To the authors' best knowledge, this represents the first demonstration of such an in situ circuit modeling approach in this field.

By jointly optimizing the reflection coefficients of all 12 calibration sources in REACH using a convolutional cost function, the method achieves an agreement of no worse than $\SI{-37}{dB}$ between measured and modeled responses. Further, by using various Bayesian inference techniques, in-depth model evaluation and comparison for components such as the loads and cables are performed. This allows for the models to be simplified where possible, as well as for their parameter biases and complexities to be better understood.

The framework is used to understand the uncertainties and sensitivities in the REACH calibration source temperature corrections. These are initially shown to exhibit large uncertainties on the order of $1$ to $\SI{2}{K}$ for measured data of the hot source, caused by parameter degeneracies, reflection coefficient biases, measurement noise, and potentially errors in the model. However, by using posteriors fit on individual components as updated priors, these uncertainties can be reduced. A simulated setup shows that, given a perfect model, this approach is capable of making precise predictions with uncertainties on the order of tens of mK while still catering for system drift. After applying these techniques to the measured data, uncertainties of around $\SI{75}{mK}$ are achieved. However, a full joint fit on this same data is found to have biases in the temperature predictions, which requires further investigation.

The techniques provide a strong foundation for future work. While this study demonstrates the application of phases 1 and 2 of the pipeline (see Fig.~\ref{fig:pipeline}), further effort is required to integrate the proposed methodologies with the calibration pipeline in phase 3. Such work should focus on quantifying the effect of these corrections on the residuals of the final calibrated antenna temperature. Additionally, the extension of the circuit models with data-driven approaches, such as Gaussian processes, warrants further investigation. In this regard, simulated data could help assess how model misspecification influences bias in the parameter estimates and temperature predictions. Finally, applying these techniques to other monitoring and correction tasks, such as for REACH's antenna, remains an open avenue for exploration.

The proposed framework enables accurate, real-time system predictions from limited measurements. This is crucial for projects such as REACH, where continuous monitoring and calibration are based solely on reflection coefficient data. By employing the described techniques, the drift in non-idealities over time can be accounted for. Ultimately, this provides a pathway towards enhanced instrumental precision and fidelity, for next-generation global 21 cm radio experiments, and radiometer calibration as a whole.

\section*{Data and Code Availability}
The data and code that supported the findings of this study are available from the corresponding author upon reasonable request.

\section*{Acknowledgments}
The research was supported by the South African Radio Astronomy Observatory, which is a facility of the National Research Foundation, an agency of the Department of Science and Technology (Grant Number: 75322). The authors would also like to thank the Kavli Foundation, and the Science and Technology Facilities Council, grant number EP/Y02916X1/1, for supporting the REACH project.

\clearpage
\appendix{}

\vspace{-20pt}

\begin{table}[!htb]
    \centering
    \setlength{\abovecaptionskip}{2pt} 
    \setlength{\belowcaptionskip}{10pt} 
    
    \caption{Load Priors}
    \label{tab:load_priors}
    \begin{tabular}{p{1.3cm}p{3.4cm}p{4.2cm}p{4.2cm}p{3.3cm}}
        \hline
        \textbf{Param.} & \textbf{Components} & \textbf{Prior (informative)} & \textbf{Prior (uninformative)} & \textbf{Unit} \\  \hline
        $R$ & Resistive & PercNormal($R_0$, $5$) & Uniform(1, 1000) & $\SI{}{\ohm}$  \\ 

        $R$ & Open & Uniform($1$, $100$) & Uniform($1$, $100$) & $\SI{}{M\ohm}$  \\ 

        $R$ & Short & Uniform(0, 10) & Uniform(0, 10) & $\SI{}{\ohm}$  \\ 
        
        $C_1$ & All & Uniform(0, 10) & Uniform(0, 100) & $\SI{}{pF}$ \\ 
        $L$ & All & Uniform(0, 10) & Uniform(0, 100) & $\SI{}{nH}$ \\ 
        $C_2$ & All & Uniform(0, 10) & Uniform(0, 100) & $\SI{}{pF}$ \\ 
        $z_n$ & All & PercNormal(50, 5) & Uniform(1, 100) & $\SI{}{\ohm}$  \\ 
                
        $\ell$ & Custom, open, short & PercNormal(37.5, 10) & Uniform(1, 100) & $\SI{}{mm}$  \\ 
        
        $\ell$ & Hot & PercNormal(23, 10) & Uniform(1, 100) & $\SI{}{mm}$  \\ 
        
        $\ell$ & Cold & PercNormal(8, 10) & Uniform(1, 100) & $\SI{}{mm}$  \\         

        $k_1$ & All & Uniform(0, 50) & Uniform(0, 100) & $\SI{}{dB}/(\SI{100}{m}\,\cdot\sqrt{\text{MHz}})$ \\ \hline        
    \end{tabular}
    
    \caption{Semi-rigid Cable Priors}
    \label{tab:semirigid_priors}
    \begin{tabular}{p{1.3cm}p{3.4cm}p{4.2cm}p{4.2cm}p{3.3cm}}
        \hline
        \textbf{Param.} & \textbf{Component(s)} & \textbf{Prior (informative)} & \textbf{Prior (uninformative)} & \textbf{Unit} \\  \hline
        $z_n$ & All & PercNormal(50, 5) & Uniform(1, 100) & $\SI{}{\ohm}$  \\ 
        
        $\ell$ & sr\_mtsj1 & PercNormal(127, 5) & Uniform(10, 1000) & $\SI{}{mm}$ \\ 
        
        $\ell$ & sr\_mtsj2 & PercNormal(101.6, 5) & Uniform(10, 1000) & $\SI{}{mm}$ \\ 
        
        $\ell$ & sr\_ms1j2 & PercNormal(100, 5) & Uniform(10, 1000) & $\SI{}{mm}$ \\ 

        $\epsilon_r$ & All & PercNormal(2.041, 5) & Uniform(1, 10) & -  \\ 
        
        $k_1$ & sr\_ms1j2 & Uniform(1.6, 2.6) & Uniform(0, 100) & $\SI{}{dB}/(\SI{100}{m}\,\cdot\sqrt{\text{MHz}})$ \\         
        $k_1$ & sr\_mtsj1, sr\_mtsj2 & Uniform(1, 3) & Uniform(0, 100) & $\SI{}{dB}/(\SI{100}{m}\,\cdot\sqrt{\text{MHz}})$ \\         
        
        $k_2$ & sr\_ms1j2 & Uniform(0.0025, 0.0075) & Uniform(0, 1) & $\SI{}{dB}/(\SI{100}{m}\,\cdot\text{MHz})$ \\ 
        $k_2$ & sr\_mtsj1, sr\_mtsj2 & Uniform(0.001, 0.01) & Uniform(0, 1) & $\SI{}{dB}/(\SI{100}{m}\,\cdot\text{MHz})$ \\ \hline        
    \end{tabular}

    \centering
    \label{tab:flexible_priors}
    \caption{Flexible Cable Priors}
    \begin{tabular}{p{1.3cm}p{3.4cm}p{4.2cm}p{4.2cm}p{3.3cm}}
        \hline
        \textbf{Param.} & \textbf{Component(s)} & \textbf{Prior (informative)} & \textbf{Prior (uninformative)} & \textbf{Unit} \\  \hline
        $a$ & All & PercNormal(1.12, 5) & Uniform(0.1, 10) & $\SI{}{mm}$  \\ 
        $b$ & All & PercNormal(3.075, 5) & Uniform(0.1, 10) & $\SI{}{mm}$  \\ 
        $\ell$ & c2 & PercNormal(2, 5) & Uniform(0.1, 100) & $\SI{}{m}$ \\ 
        $\ell$ & c10 & PercNormal(10, 5) & Uniform(0.1, 100) & $\SI{}{m}$ \\ 
        $\sigma_c$ & All & PercNormal(5.95, 5) & Uniform($1$, $100$) & $\SI{}{S/nm}$  \\ 
        $\epsilon_{r_0}$ & All & PercNormal($1.452$, 5) &  Uniform(1, 10) & - \\ 
        $\tan\delta_{0}$ & All & Uniform($0.0001$, $0.001$) &  Uniform(0, 0.1) & - \\ \hline
        
    \end{tabular}
    
    \centering
    \caption{Switch Priors}
    \begin{tabular}{p{1.3cm}p{3.4cm}p{4.2cm}p{4.2cm}p{3.3cm}}
        \hline
        \textbf{Param.} & \textbf{Component(s)} & \textbf{Prior (informative)} & \textbf{Prior (uninformative)} & \textbf{Unit} \\  \hline
        $z_n$ & All & Uniform(40, 60) & Uniform(1, 100) & $\SI{}{\ohm}$  \\         
        
        $\ell$ & ms1, ms3, ms4 & Uniform(30, 70) & Uniform(1, 1000) & $\SI{}{mm}$  \\ 
        $\ell$ & mts & Uniform(10, 100) & Uniform(1, 1000) & $\SI{}{mm}$  \\ 
        $k_1$ & All & Uniform(0, 50) & Uniform(0, 100) & $\SI{}{dB}/(\SI{100}{m}\,\cdot\sqrt{\text{MHz}})$ \\ \hline
        
    \end{tabular}
    \label{tab:switch_priors}

\caption*{Priors for parameters used in this work. Notation used is Uniform(low, high) and PercNormal($\mu$, $200 \frac{\sigma}{\mu}$) = Normal($\mu$, $\sigma$). $R_0$ represents the nominal resistance value as in Fig.~\ref{fig:reach_receiver}, with $R_0 = 50$ for the matched loads. Semi-rigid priors are written in terms of $\epsilon_r$ instead of $v_f$. $k_1$ and $k_2$ are set relatively wide for the semi-rigids due to the difficulty in obtaining a clear or reliable estimate from their datasheets.}
\end{table}

\newpage
\newpage

\bibliographystyle{ws-jai}
\bibliography{references}

\end{document}